\documentclass[acmsmall]{acmart}

\usepackage{acronym} 
\usepackage{algorithmic}
\usepackage{booktabs}
\usepackage{clrscode3e}
\usepackage{hyperref}
\usepackage{multirow}
\usepackage{graphicx}
\usepackage{diagbox}
\usepackage{subfigure}
\usepackage{textcomp}
\usepackage{xcolor}
\usepackage{tabularx}
\usepackage{threeparttable}
\usepackage{adjustbox}
\usepackage{tabularx}
\usepackage[linesnumbered,ruled,vlined]{algorithm2e}

\acrodef{RL}{Reinforcement Learning}
\acrodef{SCSR}{Shared-account Cross-domain Sequential Recommendation}
\acrodef{RL-ISN}{\textbf{R}einforcement \textbf{L}earning-enhanced \textbf{I}nformation \textbf{S}haring \textbf{N}etwork}
\acrodef{ISN}{Information Sharing Network}
\acrodef{CR}{Cross-domain Recommendation}
\acrodef{CSR}{Cross-domain Sequential Recommendation}
\acrodef{NCR}{Non-overlapping Cross-domain Recommendation}
\acrodef{NCSR}{Non-overlapping Cross-domain Sequential Recommendation}
\acrodef{MDP}{Markov Decision Process}
\acrodef{RL-DF}{Reinforcement learning-enhanced Domain Filter}
\acrodef{GRU}{Gated Recurrent Unit}
\acrodef{MLP}{Multi-Layer Perceptron}
\acrodef{SAM}{Shared-Account Modeling}
\acrodef{DA-GCN}{Domain-Aware Graph Convolutional Network}
\acrodef{BCR}{Basic Cross-domain Recommender}
\acrodef{UIN}{User Identification Network}
\acrodef{GNN}{Graph Neural Network}
\acrodef{GCN}{Graph Convolutional Network}
\acrodef{GCNs}{Graph Convolutional Networks}
\acrodef{SCRM}{Self-attention-based Cross-domain Recommendation Machine}
\acrodef{SR}{Sequential Recommendation}
\acrodef{RNNs}{Recurrent Neural Networks}
\acrodef{RNN}{Recurrent Neural Network}
\acrodef{CNN}{Convolutional Neural Network}
\acrodef{MRR}{Mean Reciprocal Rank}
\acrodef{CF}{Collaborative Filtering}
\acrodef{SA}{Self-Attention}
\acrodef{CDR}{Cross-domain Recommendation}
\acrodef{DA-GCN}{Domain-Aware Graph Convolutional Network}
\acrodef{TiDA-GCN}{Time Interval-enhanced Domain-Aware Graph Convolutional Network}
\acrodef{CDS}{Cross-Domain Sequence}
\acrodef{SAR}{Shared-Account Recommendation}
\acrodef{LLM}{Large language model}
\acrodef{MH}{Multi-head Attention}
\acrodef{TCPLP}{Text-enhanced Co-attention Prompt Learning Paradigm}
\acrodef{PLM}{Pre-trained language model}
\acrodef{MLM}{Masked Language Model}
\acrodef{CAN}{Co-Attention Network}
\acrodef{MCR}{Many-to-one Cross-domain Recommendation}
\acrodef{NMCR}{Non-overlapping Many-to-one Cross-domain Recommendation}
\acrodef{MNCSR}{Many-to-one Non-overlapping Cross-domain Sequential Recommendation}

\AtBeginDocument{%
  \providecommand\BibTeX{{%
    \normalfont B\kern-0.5em{\scshape i\kern-0.25em b}\kern-0.8em\TeX}}}

\setcopyright{acmcopyright}
\copyrightyear{2024}
\acmYear{2024}
\acmDOI{XXXXXXX.XXXXXXX}

\acmJournal{TOIS}
\acmVolume{00}
\acmNumber{0}
\acmArticle{000}
\acmMonth{0}

\begin{document}

\title{Semantic-enhanced Co-attention Prompt Learning for Non-overlapping Cross-Domain Recommendation}


\author{Lei Guo}
\authornote{Both authors contributed equally to this research.}
\email{leiguo.cs@gmail.com}
\affiliation{
  \institution{Shandong Normal University}
  \city{Jinan}
  \state{Shandong}
  \country{China}
  \postcode{250358}
}

\author{Chenlong Song}
\authornotemark[1]
\email{songchenlong0218@outlook.com}
\affiliation{
  \institution{Shandong Normal University}
  \city{Jinan}
  \state{Shandong}
  \country{China}
  \postcode{250358}
}

\author{Feng Guo}
\email{fengguo@lcu.edu.cn}
\affiliation{
  \institution{Liaocheng University}
  \city{Liaocheng}
  \state{Shandong}
  \country{China}
  \postcode{252000}
}

\author{Xiaohui Han}
\authornote{Corresponding Author.}
\email{xiaohhan@gmail.com}
\affiliation{
  \institution{Key Laboratory of Computing Power Network and Information Security, Ministry of Education, Qilu University of Technology (Shandong Academy of Sciences)}
  \city{Jinan}
  \state{Shandong}
  \country{China}
  \postcode{250014}
}
\affiliation{
  \institution{Shandong Provincial Key Laboratory of Industrial Network and Information System Security, Shandong Fundamental Research Center for Computer Science}
  \city{Jinan}
  \state{Shandong}
  \country{China}
  \postcode{250014}
}

\author{Xiaojun Chang}
\email{cxj273@gmail.com}
\affiliation{
  \institution{Shandong Normal University}
  \city{Jinan}
  \state{Shandong}
  \country{China}
  \postcode{250358}
}

\author{Lei Zhu}
\authornotemark[2]
\email{leizhu0608@gmail.com}
\affiliation{
  \institution{Tongji University}
  \city{Shanghai}
  \country{China}
  \postcode{200092}
}

\renewcommand{\shortauthors}{Guo, and Song et al.}

\begin{abstract}
Non-overlapping Cross-domain Sequential Recommendation (NCSR) is the task that focuses on domain knowledge transfer without overlapping entities. Compared with traditional Cross-domain Sequential Recommendation (CSR), NCSR poses several challenges: 1) NCSR methods often rely on explicit item IDs, overlooking semantic information among entities. 2) Existing CSR mainly relies on domain alignment for knowledge transfer, risking semantic loss during alignment. 3) Most previous studies do not consider the many-to-one characteristic, which is challenging because of the utilization of multiple source domains.
Given the above challenges, we introduce the prompt learning technique for Many-to-one Non-overlapping Cross-domain Sequential Recommendation (MNCSR) and propose a Text-enhanced Co-attention Prompt Learning Paradigm (TCPLP).
Specifically, we capture semantic meanings by representing items through text rather than IDs, leveraging natural language universality to facilitate cross-domain knowledge transfer.
Unlike prior works that need to conduct domain alignment, we directly learn transferable domain information, where two types of prompts, i.e., domain-shared and domain-specific prompts, are devised, with a co-attention-based network for prompt encoding.
Then, we develop a two-stage learning strategy, i.e., pre-train \& prompt-tuning paradigm, for domain knowledge pre-learning and transferring, respectively.
We conduct extensive experiments on three datasets and the experimental results demonstrate the superiority of our TCPLP.
Our source codes have been publicly released\footnote{https://github.com/songchenlong/TCPLP}.
\end{abstract}



\maketitle

\section{Introduction}
\textbf{\ac{SR}}~\cite{SASRec,GRU4Rec,BERT4Rec,SR-FPMC,SR_NARM,SR_Core}, as one of the main research tasks in recommendation systems, has been widely applied in various application platforms, such as Taobao.COM and JD.COM.
Different from rating-based recommenders, \ac{SR} mainly focuses on understanding users' sequential behavior for next item prediction, which is more relevant to practical applications.
However, since most users will only interact with limited items, the \ac{SR} method faces serious data sparse problems due to the lack of sufficient user behavior.
To overcome this challenge, researchers have resorted to \textbf{\ac{CSR}} for better recommendations~\cite{CSR_RL-ISN,DAGCN,PINET,GNN_DDGHM,MIFN}. \ac{CSR} is a sequential recommendation task that leverages rich information from other domains to improve the performance of the target domain, thus alleviating the performance decrease attributed to data sparsity.

\begin{figure}
    \centering
    \includegraphics[width=14cm]{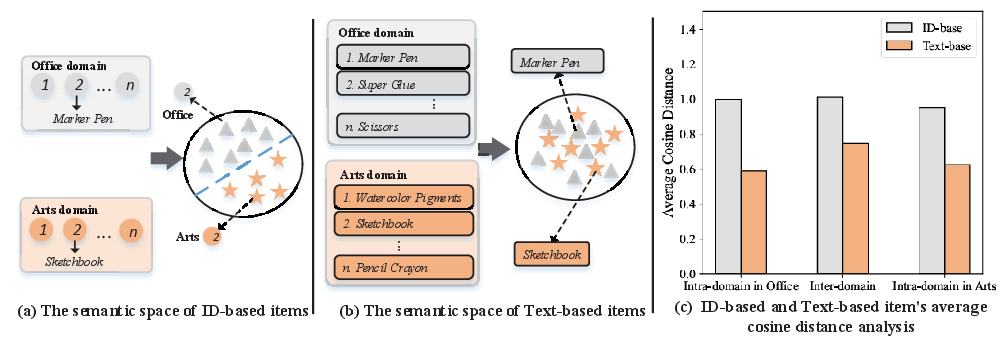}
    \caption{The semantic space differences among different forms of items in various domains. (a) illustrates the different semantic spaces of ID-based items in different domains. (b) illustrates that the text-based item can ensure that items in different domains are in the same semantic space by exploiting the generality of text in different domains. (c) compares the average cosine distances of ID-based and Text-based item representations from MCRPL and from TCPLP intra-domain and inter-domain in the \textit{Office-Arts} dataset.
    }
    \label{fig:description}
\end{figure}

In \ac{CSR} scenarios~\cite{PINET,DAGCN,overlap_seq_PSJ-Net,CSR_RL-ISN,TiDA-GCN}, different domains can be connected through overlapping users and items, which serve as bridges for knowledge transfer. Overlapping users refer to those with behavioral records in different domains, while overlapping items refer to the same items appearing in different domains. For example, users who interact with both the book domain and the movie domain can be regarded as overlapping users between the domains. Similarly, items such as "iPhone 15" may appear on both Taobao.COM and JD.COM at the same time, and these same items that appear in different domains can be regarded as overlapping items between the domains. In this scenario, shared entities (i.e., users or items) enable domain alignment and facilitate knowledge transfer between domains.
With overlapping entities as domain bridges, domain knowledge can be transferred from the source domain to the target domain for cross-domain recommendation.
However, assuming overlapping entities between domains is unrealistic in many scenarios.
In fact, many of them are totally non-overlapping. For example, in electronic news platforms and automobile news platforms, items are entirely independent. Moreover, due to the privacy requirements of the platforms, real user information is not known across platforms, making the use of overlapping entities impractical in this scenario.
Due to the lack of shared entities to bridge domains, conducting domain knowledge transfer in these scenarios is more challenging.
For solving \textbf{\ac{NCSR}}, some recent studies~\cite{NCR_TDAR,NCR_IESRec,DA-DAN} make efforts by utilizing domain adaptation techniques for domain alignment. However, since the semantic and domain information are often intertwined, directly aligning the source domain to the target domain may diminish the semantic information within the sequential behaviors.
Additionally, most current cross-domain recommendation methods~\cite{PINET,DAGCN,soft_prompt_MCRPL,soft_prompt_PLCR,UniCDR,overlap_seq_PSJ-Net} represent items only based on IDs. But as items in different domains are in different semantic spaces, as illustrated in Fig.~\ref{fig:description} (a), the complete use of atomic item IDs is unsuitable for cross-domain scenarios.
The incapability of modeling the semantic information enables the ineffectiveness of these ID-based methods due to the lack of sufficient user data, especially in cold-start domains.

In this study, we focus on the emerging yet challenging \textbf{\ac{MNCSR}} task.
The `many-to-one' characteristic refers to the task that tends to transfer knowledge from multiple source domains to enhance a single target domain in cross-domain recommendation tasks, where the target domain often has sparse data issues and source domains have relatively dense data collections.
We consider the many-to-one characteristic since: 1) It is a general practice for platforms to improve the target domain by leveraging more than one domain.
Using multiple source domains can further increase the effectiveness of information fusion due to richer information.
2) Utilizing multiple source domains for cross-domain recommendation tasks provides potential opportunities to address distribution differences and shift issues between source and target domains, since multiple source domains can offer a wider range of domain variation patterns, thus improving the robustness of recommendation systems.
In our non-overlapping scenario, this task becomes more challenging, because of the absence of shared users and items among domains.
Two challenges remain for conducting \ac{MNCSR}:
1) Due to the absence of any overlapping entities between domains, interaction and information sharing between domains becomes difficult.
2) Since semantic and domain information are often tightly intertwined, direct alignment of source and target domains may trigger the loss of semantic information. It remains a challenge to conduct effective knowledge transfer while ensuring that semantic information is not lost.
There are several works that have paid attention to this direction. For example, Liu et al.~\cite{soft_prompt_MCRPL} use prompt learning to address \ac{MNCSR}, and Zhang et al.~\cite{nonoverlap_many_to_one_1} utilize domain adaptation techniques to mitigate domain shift.
But they primarily use ID-based methods and fail to capture semantic information, thereby impacting the recommendation performance.

Inspired by recent successful large language models~\cite{LLM_T5,LLM_Bert,LLM_longformer} and the universality of natural language texts, we resort to the semantic texts for domain bridging in the \ac{MNCSR} scenario.
Although items in different domains are represented in different atomic IDs (they are usually in different feature spaces in traditional ID-based methods), their content information may disclose the same semantic meaning as the universality of the natural language.
An illustrated example of our text-based item representation is shown in Fig.~\ref{fig:description} (b).
We illustrate this experimentally by using cosine distance as the metric to evaluate the alignment of Text-based and ID-based item representations both within each domain and across domains. The experimental results are shown in Fig.~\ref{fig:description} (c), where we find that text-based item representations exhibit significantly lower inter-domain distances and consistent intra-domain distances. This demonstrates that text-based item representations are better aligned across domain semantics while retaining domain-specific nuances.
Specifically, we propose a \textbf{\ac{TCPLP}} for \ac{MNCSR}, which mainly consists of a text sequence encoder, a co-attention-based prompt encoder, and a two-stage pre-train \& prompt tuning strategy. 
More concretely, to capture the semantic meanings of items and establish connections between non-overlapping domains, we utilize the generality of natural language text and the powerful representation capabilities of the pre-trained language models to represent items and sequences.
Different from previous solutions conducting domain knowledge transfer by forcing alignment of domains, we solve it by directly learning common domain features without domain alignment.
More concretely, we present two types of prompts: domain-shared prompts and domain-specific prompts, aiming at extracting invariant features and specific features from multiple domains, respectively. 
To accurately learn the representations of these prompts, we introduce a co-attention network to capture the correlation between sequence and prompt contexts.
These pre-learned domain prompts are seen as the common knowledge of all the engaged domains.
As we represent items in natural language texts, we can easily embed them into the same semantic space.
To further adapt these pre-learned knowledge into specific domains, a prompt tuning paradigm is then developed with domain-shared prompts and co-attention network fixed.
In our pre-train \& prompt tuning learning strategy, the pre-training stage leverages sequence data from various domains, including the target domain, to train our text sequence encoder and capture prior knowledge across domains. Subsequently, the prompt-tuning stage refines domain-specific prompts exclusively within the target domain to reinforce the learning of domain-specific knowledge in the target domain.

The main contributions of this work are summarized as follows:
\begin{itemize}
\item We study \ac{MNCSR} and solve it by proposing \ac{TCPLP}, which utilizes a semantic-enhanced co-attention network for knowledge modeling and a prompt learning strategy for domain transfer.
\item We model items' semantic meanings by their texts rather than atomic IDs, and leverage the universality of natural language as a bridge to connect non-overlapping domains in the same semantic space.
\item We conduct knowledge transfer by directly learn the common features among domains instead of domain alignment, where two types of prompts, domain-shared prompts and domain-specific prompts, are designed, and a co-attention-based encoder is devised for its encoding.
\item We develop a two-stage learning paradigm, i.e., pre-training stage, and prompt tuning stage, to learn prior domain knowledge and then adapt them to the specific target domain.
\item We conduct extensive experiments on two subsets of the Amazon dataset and a cross-platform dataset to evaluate the performance of our proposed method. The experimental results demonstrate that our method outperforms state-of-the-art baselines.
\end{itemize}
\section{Related Work}
This section introduces four types of recommendation methods, namely cross-domain recommendation in overlapping scenarios, cross-domain recommendation in non-overlapping scenarios, multi-domain recommendation, prompt learning-based recommendation, and text-based recommendation, as our related works. Subsequently, we discuss the differences between our approach and the related methods.

\subsection{Cross-domain Recommendation in Overlapping Scenarios}
\textbf{\ac{CDR}} aims to leverage rich source domain information to enhance the target domain and alleviate the data sparsity issue in the target domain. Existing \ac{CDR} methods mostly rely on modeling overlapping entities. By mining the associative relationships between these overlapping entities, knowledge from the source domain can be effectively transferred to the target domain, thereby achieving the goal of \ac{CDR}. Depending on the form of data, we can divide \ac{CDR} into rating-based \ac{CDR} and sequence-based \ac{CDR}.

Rating-based \ac{CDR} methods primarily focus on users' rating data for items. Researchers often classify tasks into single-target~\cite{overlap_rating_single_1,overlap_rating_single_2,overlap_rating_single_3,overlap_rating_single_4}, dual-target~\cite{overlap_rating_dual_1,overlap_rating_dual_2,overlap_rating_dual_3,overlap_rating_dual_4,overlap_rating_dual_5,overlap_rating_dual_6}, and multi-target~\cite{overlap_rating_multi_1,overlap_rating_multi_2} tasks. For example, Jiang et al.~\cite{overlap_rating_single_1} utilize low-rank sparse decomposition to capture common rating patterns in relevant domains while segmenting specific domain patterns to achieve heterogeneous recommendation in the target domain. However, this approach only enhances the accuracy of target domain recommendation through the source domain and does not simultaneously improve the recommendation performance in both domains. Li et al.~\cite{overlap_rating_dual_1} employ the concept of deep dual transfer learning by learning latent orthogonal mappings across domains and utilizing user preferences from all domains to provide cross-domain recommendation, aiming to simultaneously enhance the recommendation performance in both domains. Subsequently, researchers extended this to multiple domains to simultaneously enhance the performance across all domains. Cui et al.~\cite{overlap_rating_multi_1} introduce a shared structure to model information from multiple domains (such as graphs) aiming to improve the performance across all domains.

Sequence-based \ac{CDR} methods involve mining user behavior sequences in both the source and target domains to infer users' interest evolution and behavior pattern changes, thereby enabling \ac{CSR}. \ac{CSR} has seen numerous innovative research works~\cite{PINET,overlap_seq_PSJ-Net,CSR_RL-ISN,DAGCN,GNN_DDGHM,GNN_C2DSR,overlap_seq_DASL, MIFN}. For example, $\pi$-Net~\cite{PINET} is encoded by an RNN-based sequence encoder and utilizes a parallel information-sharing network to transmit relevant data across domains. With the powerful modeling capability of graph neural networks, Guo et al.~\cite{DAGCN}, Zheng et al.~\cite{GNN_DDGHM}, and Cao et al.~\cite{GNN_C2DSR} employ graph networks to model sequence information, mining behavioral relationships between sequences, and further transferring relevant information. Ma et al.~\cite{MIFN} simultaneously consider behavioral information flow and cross-domain knowledge flow through a behavior transfer unit and a knowledge transfer unit. Guo et al.~\cite{CSR_RL-ISN} utilize \ac{RL} to enhance user-specific account representations and mitigate the dilution of domain information, achieving \ac{SCSR}. However, these methods capture information and transfer it through overlapping entities as bridges between domains.

\subsection{Cross-domain Recommendation in Non-overlapping Scenarios}
In real-world scenarios, the existence of overlapping entities across different domains does not meet practical requirements. Therefore, achieving cross-domain recommendation in non-overlapping scenarios becomes highly important. However, due to the lack of direct connections between domains in non-overlapping scenarios, it poses greater challenges. Currently, \textbf{\ac{NCR}} can be categorized by task type into single-target~\cite{nonoverlap_single_1,nonoverlap_single_2,nonoverlap_single_3,nonoverlap_single_4,nonoverlap_single_5,nonoverlap_single_6,NCR_IESRec}, dual-target~\cite{nonoverlap_dual_1,nonoverlap_dual_2}, multi-target~\cite{nonoverlap_mult_1}, one-to-many~\cite{nonoverlap_one_to_many_1}, and many-to-one~\cite{nonoverlap_many_to_one_1,nonoverlap_many_to_one_2,soft_prompt_MCRPL} tasks.

For example, Perera et al.~\cite{nonoverlap_single_3} achieve this by learning a mapping from the preference manifold of the target network to the preference manifold of the source network, thereby synthesizing the preferences of non-overlapping users from the source network. The generated preferences can then be utilized to enhance recommendations for non-overlapping users. The method described above does not take into account improving the performance of the source domain. Zhang et al.~\cite{nonoverlap_dual_1} employ dual adversarial learning, utilizing domain discriminators to train a shared encoder to maximize the alignment between the latent feature spaces of users and items across the source and target domains. Consequently, it enables better recommendations for both domains in non-overlapping scenarios. Li et al.~\cite{nonoverlap_mult_1} handle multi-objective non-overlapping tasks by establishing correlations between multiple rating matrices through finding shared implicit clustering-level rating matrices. Krishnan et al.~\cite{nonoverlap_one_to_many_1} utilize domain-invariant components shared between dense and sparse domains to guide neural collaborative filtering, thereby leveraging representations of users and items learned in multiple sparse domains to improve performance in a dense domain. However, knowledge extracted solely from one source domain may not be sufficient, as there may be useful knowledge available from many other source domains. Zhang et al.~\cite{nonoverlap_many_to_one_1} extract group-level knowledge from multiple source domains to improve recommendations in a sparse target domain.

\subsection{Multi-domain Recommendation}
In this study, `many-to-one' refers to the task that transfers knowledge from multiple source domains to enhance a single target domain, where the target usually has sparser data collections.
The multi-domain recommendation task aims to utilize knowledge from other domains to improve the overall performance of the multi-domain, whose optimization objective is to optimize all the interacted domains.

The existing methods on multi-domain recommendation tasks can be categorized into multi-task learning~\cite{mtr-apt, mtr-1, mtr-star, mtr-2} and dynamic parameter tuning~\cite{multi-pepnet, multi-adasparse, multi-m2m}.
Multi-task learning mainly focuses on improving its multi-domain performance simultaneously by considering each domain as one optimization objective through a multi-task optimization approach. Sheng et al.~\cite{mtr-star} propose a star topology consisting of shared center parameters and domain-specific parameters to satisfy the click-through rate prediction of one model for different domains. Hao et al.~\cite{mtr-apt} propose a generator-discriminator framework utilizing a generator-discriminator framework to learn user interests in all domains to solve the multi-domain recommendation task.

Dynamic parameter tuning is mainly utilized to construct a global shared network while dynamically tuning specific module parameters based on domain-specific samples. For example, Yang et al.~\cite{multi-adasparse} prune redundant neurons by adaptively learning sparse structures for each domain and using domain-aware neuron-level weighting factors. Zhang et al.~\cite{multi-m2m} propose a multi-scenario, multi-task meta-learning model that introduces meta-units to merge scenario knowledge by generating weights for the backbone model.
Chang et al.~\cite{multi-pepnet} take personalized a priori information as input and dynamically adjust the size of the bottom embedding and top DNN hidden units through a gating mechanism.

\subsection{Prompt Learning-based recommendation}
With the rapid development of pre-trained language models, prompt learning becomes a highly regarded field. Prompt learning guides models to generate specific types of outputs by using concise prompt statements without significantly altering the structure and parameters of pre-trained language models. Currently, prompts are mainly divided into two categories: discrete prompts and continuous prompts. Discrete prompts are typically manually designed text used to directly prompt the model for the desired task or objective. In contrast, continuous prompts usually consist of learnable embeddings, offering a more flexible representation that allows the model to learn how to generate task-related outputs from the data. Recent studies combine prompt learning with recommendation systems to improve their performance and effectiveness. We classify them into two categories: discrete prompts based recommendation methods~\cite{hard_prompt_p5,hard_prompt_genrec,hard_prompt_gpt4rec,hard_prompt_Prompt4NR,hard_prompt_PEPLER} and continuous prompts based recommendation methods~\cite{soft_prompt_ppr,soft_prompt_plate,soft_prompt_PLCR,soft_prompt_MOP,soft_prompt_MCRPL}.

Discrete prompts based recommendation methods primarily utilize manually designed prompt statements that reflect users' interests and preferences to fine-tune large language models, enabling the generation of personalized recommendation results based on users' discrete prompt information. For example, Geng et al.~\cite{hard_prompt_p5}, Ji et al.~\cite{hard_prompt_genrec}, Li et al.~\cite{hard_prompt_gpt4rec}, Zhang et al.~\cite{hard_prompt_Prompt4NR}, and Li et al.~\cite{hard_prompt_PEPLER} utilize discrete prompt templates composed of natural language to fine-tune large language models for recommendation purposes.

Continuous prompt-based recommendation methods can automatically learn some specific knowledge through prompt embedding, which can better guide downstream tasks. For example, Wu et al.~\cite{soft_prompt_ppr} generate personalized continuous prompts based on user preferences and achieve comprehensive training through prompt-based contrastive learning, effectively addressing the challenge of cold-start recommendation.
Hao et al.~\cite{soft_prompt_MOP} propose a motif-based prompt learning framework, introducing motif-based shared embeddings to encapsulate generalized domain knowledge and aligning training objectives across domains.

\subsection{Text-based Recommendation}
Recently, inspired by the success of large language models and the universality of natural language text, some works begin utilizing natural language text instead of traditional IDs to represent items in recommendation.
We can categorize existing research into two types: text representation-based methods~\cite{text_recformer,text_UniSRec,text_LLM-REC,hard_prompt_gpt4rec,text_TALLRec,text_llamaRec,text_llara,text_llmseqsim,text_ida-sr} and code representation-based methods~\cite{text_Tiger, text_VQRec}. 

Text representation-based method, by using the language model encoder to encode the item text representation. For example, Li et al.~\cite{text_recformer} and Tang et al.~\cite{text_LLM-REC} utilize text to represent items and encode sequences using language models. 

Code representation-based methods use the item text to generate discrete ID, and use discrete ID to process the recommendation task. For example, Hou et al.~\cite{text_UniSRec} design a lightweight item encoding architecture based on parameter whitening and a mixture-of-experts enhanced adaptor to learn a generic item representation.
Hou et al.~\cite{text_VQRec} map item text to discrete indices and employ enhanced contrastive pre-training and cross-domain fine-tuning methods, effectively addressing issues of over-reliance on text features and exaggeration of domain gaps.
Rajput et al.~\cite{text_Tiger} utilize a hierarchical method to generate semantic IDs from item text and directly generate predictions using an autoregressive approach.

\subsection{Differences}
Our work is different from existing studies in the following aspects: 1) Our \ac{TCPLP} method is different from the MCRPL~\cite{soft_prompt_MCRPL} and PLCR~\cite{soft_prompt_PLCR} methods in \ac{NCSR}, as they only exploit the atomic IDs to represent items and do not utilize the textual information to capture the semantic meanings of items, while ours leverage the university of natural languages to bridge the distributional differences among non-overlapping domains. 2) Our work differs from other prompt-based cross-domain recommendation methods, such as Wang et al.~\cite{soft_prompt_plate}, which utilizes prompt enhancement for multi-scenario recommendation. However, they do not consider sequence information or utilize textual information. Moreover, we devise a co-attention-based network for prompt encoding.
3) Our method is also different from other text-based \ac{CDR} methods that they do not consider enhancing the capture of domain-invariant features and domain-specific features, lacking effective knowledge transfer between domains. For example, Hou et al.~\cite{text_VQRec} and Rajput et al.~\cite{text_Tiger} transform item texts from different domains into discrete codes, enabling items from different domains to be discretely encoded based on the similarity of their text semantics, where items with similar text semantics have more similar discrete codes, resulting in closer distances in the feature space. However, they do not consider transferring some shared knowledge to compensate for domain differences.
4) Our TCPLP and PFCR~\cite{PFCR} are different in semantic representation and knowledge transfer mechanisms, although they both connect different domains through text. 
Firstly, PFCR employs vector quantization to transform item text into multiple discrete index codes, which are then used to look up the code embedding table to derive the item representation. However, the discrete codes generated from the original item text through quantization can only reflect the approximate semantics within each subspace, and this process inevitably leads to some loss of the semantics of the original text. TCPLP directly maps the text to a continuous dense vector as the item representation using a pre-trained language model, avoiding semantic loss caused by discrete encoding.
In addition, TCPLP and PFCR employ different methods in transferring knowledge between non-overlapping domains. PFCR primarily relies on a shared codebook for item-level cross-domain information sharing on the basis of the discrete item codes.
Different from PFCR, our TCPLP discards the traditional domain alignment strategy and adopts a sequence-level co-attention network to realize cross-domain knowledge capture. The network can establish dynamic associations between sequence representations and prompt contexts, enabling different types of prompts to have different impacts on the sequence representations for better embedding of a priori domain knowledge. During the pre-training stage, the co-attention network is used to learn the cluster-level shared features from different domains, so as to achieve sequence-level cross-domain knowledge sharing.

\section{METHODOLOGY}
In this section, we first provide the preliminaries of this work and 
then provide the details of we develop the \ac{TCPLP} method for the \ac{MNCSR} task.

\begin{figure*}[!t]
    \centering
    \includegraphics[width=12cm]{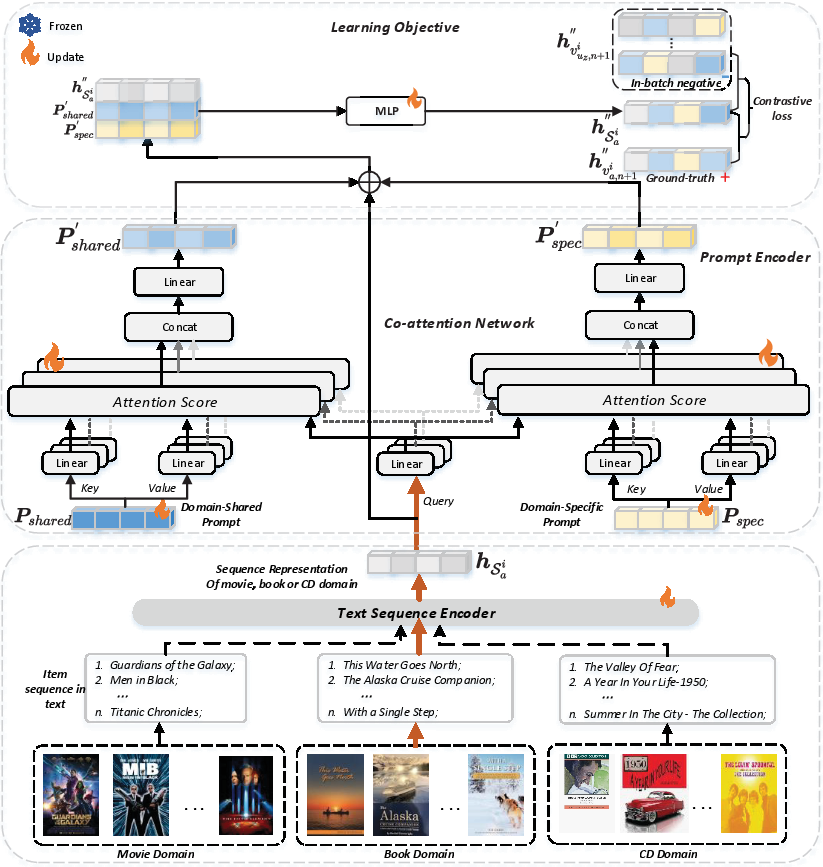}
    \caption{
    The system architecture of \ac{TCPLP} in the pre-training stage, which mainly comprises three components: the text sequence encoder, prompt encoder, and the learning framework. At the same time, we utilize multiple domain text sequences as input for pre-training and prior knowledge learning to train the text sequence encoder, prompt encoder, and MLP network.}
    \label{fig:overview}
\end{figure*}

\subsection{Preliminaries}
Assuming we have $N$ domains (e.g., $\mathcal{D}=\left \{ {D}_{1},{D}_{2},..., {D}_{T}, ...,{D}_{N}  \right \} $), where $D_T$ is the target domain and the rest are source domains. Let $\mathcal{U}^i =\left \{ {u}^i_{1},{u}^i_{2},..., {u}^i_{a}, ...,{u}^i_{n_i}  \right \} $, $\mathcal{V}^i =\left \{ {v}^i_{1},{v}^i_{2},..., {v}^i_{b}, ..., {v}^i_{m_i}  \right \} $ denote the user set and item set in domain $D_i$, where $n_i$ and $m_i$ represent the number of users and the number of items in domain ${D}_i$, respectively. 
All domains do not have any overlapping entities (i.e., user and item) among them, which means that we can only observe users' interaction behaviors on items in specific domains.

Formally, let $\mathcal{S}_{a}^{i} = \left \{ v^i_{a,1}, v^i_{a,2}, ..., v^i_{a,n} \right \} $ be the interaction sequence of user $u_a^i$ in domain ${D}_i$, where $n$ is the sequence length. For each sequence $\mathcal{S}_{a}^{i}$, we use the last item of it as the ground truth label. 
To represent items in a universal feature space and take advantage of the natural language in bridging different domains, we denote items by their descriptive titles, rather than the atomic item IDs.
The textual representation of $v^i_b$ is indicated by $\mathcal{T}^i_b = \left \{ w_1, w_2, ..., w_c, ..., w_L \right \} $, where $w_c$ is the content word in natural languages, and $L$ is the truncated length of item text. We construct a text sequence $ \left \{ \mathcal{T}^i_{a,1}, \mathcal{T}^i_{a,2}, ..., \mathcal{T}^i_{a,n}\right \}$ for user $u_a^i$ by arranging the interaction sequence $\mathcal{S}_{a}^{i}$ in sequence order. Table~\ref{tab:symbol} summarizes the symbols and their notations used.
Our target is to utilize multi-source domain information to enhance the target domain, aiming at alleviating the data sparsity issue in the target domain. Different from traditional \ac{CDR} tasks, we focus on the non-overlapping scenario and tend to solve it by devising a pre-train \& prompt tuning paradigm. The input and output of our task are defined as:

\textbf{Input:} The text sequence of user interactions $\{ \mathcal{T}^{D_i}_{a,1}, \mathcal{T}^{D_i}_{a,2}, ..., \mathcal{T}^{D_i}_{a,n} \} $ in all domains, where $D_i \in \mathcal{D}$.

\textbf{Output:} The probability $P(v_{a,n+1}^T \mid \left \{ \mathcal{T}^{T}_{a,1}, \mathcal{T}^{T}_{a,2}, ..., \mathcal{T}^{T}_{a,n} \right \})$ of recommending $v_{a,n+1}^T$ to be the interacted item by a given user $u^{T}_{a}$ in the target domain $D_T$.

\begin{table}[htbp]
  \centering
  \caption{Summary of the Utilized Symbols and Notations in this Work.}
  \scalebox{0.91}{
  \begin{tabular}{ll}
    \toprule
    \textbf{Symbols} & \textbf{Description} \\
    \midrule
    $\mathcal{D}$ & Domain set $\left \{ {D}_{1},{D}_{2},..., {D}_{T}, ...,{D}_{N}  \right \} $, where $D_{T}$ is the target domain among $N$ domain. \\
    $\mathcal{U}^{i}$ & User set $\left \{ {u}^i_{1},{u}^i_{2}, ..., {u}^i_{a}, ...,{u}^i_{n_i}  \right \} $ in domain $D_{i}$, where $n_{i}$ is the number of users in domain $D_i$.\\
    $\mathcal{V}^{i}$ & Item set $\left \{ {v}^i_{1},{v}^i_{2}, ..., {v}^i_{b}, ...,{v}^i_{m_i}  \right \} $ in domain $D_{i}$, where $m_i$ is the number of items in domain $D_i$.\\
    $\mathcal{S}^{i}_{a}$ & The interaction sequence $\left \{ v^i_{a,1}, v^i_{a,2}, ..., v^i_{a,n} \right \}$ of user $u^{i}_{a}$ in domain $D_{i}$.\\  
    $\mathcal{T}^{i}_{b}$ & The title text $\left \{ w_1, w_2, ..., w_L \right \}$ of item $v^{i}_{b}$, where $L$ is the truncation length of the text.\\  
    $\boldsymbol{P}_{shared}$ & Domain-shared prompts.\\
    $\boldsymbol{P}_{spec}$ & Domain-specific prompts.\\
    $\boldsymbol{h}_{\mathcal{S}^{i}_{a}}$ & The textual representation of user $u^{i}_{a}$ in domain $D_{i}$.\\
    $\boldsymbol{h}_{v^{i}_{b}}$ & The representation of item $v^{i}_{b}$ in domain $D_i$. \\
    $\boldsymbol{P}_{shared}^{'}$ & Enhanced domain-shared prompts through Co-attention network.\\    
    $\boldsymbol{P}_{spec}^{'}$ & Enhanced domain-specific prompts through Co-attention network.\\    
    $\boldsymbol{h}_{\mathcal{S}^{i}_{a}}^{'}$ & The representation of sequence $\mathcal{S}^{i}_{a}$ after domain information fusion.\\
    $\boldsymbol{h}_{v^{i}_{b}}^{'}$ & The representation of item $v^{i}_{b}$ after domain information fusion.\\
    $\boldsymbol{h}_{\mathcal{S}^{i}_{a}}^{''}$ & The representation of $\boldsymbol{h}_{\mathcal{S}^{i}_{a}}^{'}$ after dimensionality reduction through an MLP network.\\
    $\boldsymbol{h}_{v^{i}_{b}}^{''}$ & The representation of $\boldsymbol{h}_{v^{i}_{b}}^{'}$ after dimensionality reduction through an MLP network.\\  
    $\mathcal{L}_{pre}$ & The contrastive loss in the pre-training phase.\\ 
    $\mathcal{L}_{pt}$ & The contrastive loss in the prompt-tuning phase.\\   
    $\tau$ & Temperature coefficient for contrastive loss.\\   
    $d_W$ & The number of prompt contexts.\\   
    
    \bottomrule
  \end{tabular}
  }
  \label{tab:symbol}
\end{table}

\subsection{Overview of \ac{TCPLP}}
\textbf{Motivation.}
Due to the sparsity of users' interactions in recommender systems, models face challenges in accurately predicting user interests~\cite{Scarcity_1, Scarcity_2}.
\ac{CDR} uses information from different domains to establish a shared user representation between different domains through overlapping users or items.
A common practice of previous methods is to transfer users' invariant preference across domains by taking the overlapped users as bridges to align domain.
However, this approach is not applicable to the \ac{NCR} scenarios, since the lack of the shared users between domains.

To connect the disjoint domains, we resort to the textual information in natural languages, which could play the role of domain bridge in a general language semantic space.
To unify the item text in the same semantic space and learn the pre-knowledge of all domains, we first universally pre-learn the domain features with a co-attention network to embed the prior knowledge into prompts in the pre-training stage.
Specifically, two types of prompts, namely domain-shared prompt and domain-specific prompt, are designed for knowledge embedding and transfer. The domain-shared prompt is responsible for learning the invariant domain features shared by all the domains. The domain-specific prompt focuses on learning the unique features that are specific to each domain.
Then, to adapt the pre-learned domain knowledge to the target domain, we conduct further optimization in the prompt-tuning stage, where we freeze the pre-trained text sequence encoder and domain-shared prompts, and only tune the domain-specific prompts to meet the specific distribution of the target domain.
In our solution, prompt can be seen as a kind of cluster-level domain feature, which can be transferred without domain alignment.


\textbf{Overall Framework.}
We adopt a two-stage strategy, i.e., a pre-train \& prompt tuning paradigm, for optimization.
The system architecture of the pre-training stage is shown in Fig.~\ref{fig:overview}, which consists of a text sequence encoder and a co-attention-based prompt encoder.
The text sequence encoder is designed to embed text sequences of items in all domains, obtaining users' sequential preference in semantic space.
The prompt encoder aims to embed domain knowledge into two kinds of prompts by exploring the co-attention network, where the domain-shared prompt tends to retain the domain invariant features, while the domain-specific prompt intends to keep the domain specific features in the prompt-tuning stage.
The architecture of \ac{TCPLP} in the prompt tuning stage is shown in Fig.~\ref{fig:finetune}, which is only trained on the target domain aiming at adapting the pre-learned domain knowledge to the specific domain.
In this stage, the pre-learned text sequence encoder and domain-shared prompts are kept fixed to retain the invariant domain information, while the domain-specific prompts are further fine-tuned to fit the target domain.

\subsection{Text Sequence Encoding}
We employ texts instead of atomic IDs to represent items, since in \ac{NCSR} scenario, users and items are non-overlapped, and there is a big gap in data distributions between two domains. It is hard to directly align them without any common traits.
To connect these disjoint domains, we draw on the commonality of natural language, which can represent items in the general semantic space and find their similarity by the descriptive texts on them.
Based on this, we can model items in a unified feature space, facilitating the knowledge transfer across domains.
To enable the text sequence encoder to have profound language understanding and the generation of high-quality text representations, pre-trained language models that have been trained on large-scale text data are exploited. In this work, we exploit Longformer~\cite{LLM_longformer} and BERT~\cite{LLM_Bert} as the backbone of our text encoder and fine-tune them in an end-to-end manner within the pre-training stage, where the word embeddings in Longformer and BERT are kept fixed.

\textbf{Model Inputs.}
For each item $v_{a,j}^i$ within the given sequence $\mathcal{S}_{a}^{i} = \left \{ v_{a,1}^i, v_{a,2}^i, ..., v_{a,j}^i, ..., v_{a,n}^i \right \}$ in domain $D_i$, we denote it by its descriptive texts (title is utilized in this work), indicated by $\mathcal{T}_{a,j}^i = \left \{ w_1, w_2, ..., w_L \right \}$. We add a specific token `[CLS]' at the beginning of $\mathcal{T}_{a,1}^{i}$ (i.e., the whole text sequence of $\mathcal{S}_a^i$)~\cite{LLM_Bert, LLM_longformer, text_recformer, text_LLM-REC} and output its embedding as the representation of $\mathcal{S}_a^i$. 
The text sequence of the input $ \mathcal{S}_{a}^i$ can be formalized as:
\begin{equation}
\operatorname{Input}\left ( \mathcal{S}_{a}^i \right ) = \left \{ \left [ \text{CLS}\right ], \mathcal{T}_{a,1}^{i}, \mathcal{T}_{a,2}^{i}, ..., \mathcal{T}_{a,n}^{i}\right \}. 
\label{Input}
\end{equation}

During the pre-training stage, we simultaneously input the text sequences from all domains. This is done because we observe that the texts from different domains possess their own unique linguistic styles, vocabularies, and topics. By encoding them in a shared feature space, we can capture these commonalities and establish cross-domain language understanding capabilities within the model, enhancing the model's generalization and representation capabilities.

\textbf{Encoding via Language Models.}
We introduce two widely used pre-trained language models for sequence and item encoding: Longformer~\cite{LLM_longformer} and BERT~\cite{LLM_Bert}.
Longformer employs a global attention mechanism, combining local and global attention mechanisms to effectively handle long sequences without significantly increasing computational resources as the sequence length grows.
BERT exploits a bidirectional Transformer architecture for sentence modeling by optimizing the masked language model task.
It has significant advancements in understanding text context and generating text representations, and has been widely applied in various natural language processing tasks.
Both items and sequences are modeled by the above two language models with an item is taken as a sequence of length 1. 

Sequence modeling. For a given sequence $\mathcal{S}_a^i$ in domain $D_i$, we can learn its high dimensional representation by:
\begin{equation}
\boldsymbol{h}_{\mathcal{S}_{a}^{i}} = \operatorname{PLM}\left ( \left \{ [\text{CLS}], \mathcal{T}_{a,1}^{i}, \mathcal{T}_{a,2}^{i}, ..., \mathcal{T}_{a,n}^{i} \right \}  \right ),
\label{hs}
\end{equation}
where $\boldsymbol{h}_{\mathcal{S}_{a}^{i}}$ denotes the representation of the sequence $\mathcal{S}_a^{i} = [v_{a,1}^{i},v_{a,2}^{i},...,v_{a,n}^{i}]$. $\operatorname{PLM}(\cdot)$ is the encoder and pooler component defined in BERT and Longformer to learn the representation of text sequences.

Item embedding.
We regard item as a special sentence of length 1, and exploits the same method for its representation learning. The learning for item $v_{b}^{i}$ can be defined as:
\begin{equation}
\boldsymbol{h}_{v_{b}^{i}} = \operatorname{PLM}\left ( \left \{ [\text{CLS}], \mathcal{T}_{b}^{i} \right \}  \right ),
\label{hv}
\end{equation}
where $\boldsymbol{h}_{v_{b}^{i}}$ represents the representation of item $\boldsymbol{v}_{b}^{i}$.

\subsection{Prompt Encoding via Co-attention Network}

\textbf{Design of Prompts.}
In general, domain knowledge among domains can be categorized into domain-invariant and domain-specific information, where domain-invariant information is the features that are shared by both domains, and domain-specific information indicates the aspects that are unique in different domains.
To extract these two kinds of information from the semantic meaning of the sequences, we leverage two types of prompts, i.e., domain-shared and domain-specific prompts.
They are respectively designed for capturing domain-invariant and domain-dependent features, and learned by a co-attention network.


Let $\boldsymbol{P}_{shared} \in \mathbb{R}^{d_W \times d_V}$ and $\boldsymbol{P}_{spec} \in \mathbb{R}^{d_W \times d_V}$ denote domain-shared and domain-special prompts respectively. Where $d_W$ represents the number of prompt contexts, which determines the richness and diversity of knowledge inside the prompt.
We set $d_V$ to be consistent with the dimensionality of sequence representations for ease of computation.
Both domain-shared prompts and domain-specific prompts are learnable continuous prompts.
For the domain-shared prompts ($\boldsymbol{P}_{shared} \in \mathbb{R}^{d_W \times d_V}$), we initialize them by using random initialization.
In the pre-training stage, they are optimized using multi-domain data. Multi-domain data contain various textual descriptions from different domains, allowing the prompts to be updated to capture common semantic features shared across multiple domains.
For the domain-specific prompts ($\boldsymbol{P}_{spec} \in \mathbb{R}^{d_W \times d_V}$), we initialize them by the same random initialization approach. In the pre-training stage, we similarly optimize them using multi-domain data to ensure that they have better initial features before the prompt-tuning stage. In the prompt-tuning stage, we optimize them again using the target domain data to adapt the $\boldsymbol{P}_{spec}$ to the specific data distribution of the target domain.

Note that, we model the shared common knowledge and the domain-specific information in two stages.
In the pre-training stage, we tend to utilize $\boldsymbol{P}_{shared}$ and $\boldsymbol{P}_{spec}$ to model the pre-knowledge of all the domains, such as the common subject information in the book and movie domains.
Then, we fine-tune $\boldsymbol{P}_{spec}$ in the prompt-tuning stage with $\boldsymbol{P}_{shared}$ is kept fixed. This is to adapt $\boldsymbol{P}_{spec}$ to meet the specific data distribution of the target domain and enhance cross-domain processing by the fixed domain-invariant information.

\textbf{Co-attention Network.}
To enable different prompt contexts have various impacts to the domain semantics, we devise a co-attention network to facilitate the interaction between sequence representations and prompts, where the sequence embedding is severed as a co-guidance in a parallel attention network. 
In the co-attention network, the attention of various components of the prompts is dynamically adjusted on the basis of the sequence meanings. This adjustment enables the attention network to more accurately capture the correlations between sequences and the prompt contexts in both domain-shared and domain-specific prompts, facilitating a better embedding of prior domain knowledge.

Two parallel \ac{MH} networks~\cite{attention} are utilized to calculate the importance of the prompt contexts to the given sequence $\mathcal{S}_a^i$ in all domains.
Specifically, we treat the sequence representation $\boldsymbol{h}_{\mathcal{S}_a^{i}}$ as query ($\boldsymbol{Q}$) and the prompt contexts as keys ($\boldsymbol{K}$) and values ($\boldsymbol{V}$), where $\boldsymbol{h}_{\mathcal{S}_a^{i}}$ is shared by both attention networks. Then, we calculate attention distributions to aggregate the prompts in a weighted manner.
We denote the encoded domain-shared and domain-specific prompts as $\boldsymbol{P}_{shared}^{'}$ and $\boldsymbol{P}_{spec}^{'}$, respectively.
The encoding network can be formulated as:
\begin{equation}
\boldsymbol{P}_{shared}^{'} = \operatorname{MH}\left ( \boldsymbol{h}_{\mathcal{S}_{a}^{i}}\boldsymbol{W}^{Q}_1, \boldsymbol{P}_{shared}\boldsymbol{W}^{K}_1, \boldsymbol{P}_{shared}\boldsymbol{W}^{V}_1 \right ),
\label{p_share}
\end{equation}

\begin{equation}
\boldsymbol{P}_{spec}^{'} = \operatorname{MH}\left ( \boldsymbol{h}_{\mathcal{S}_a^{i}}\boldsymbol{W}^{Q}_2, \boldsymbol{P}_{spec}\boldsymbol{W}^{K}_2, \boldsymbol{P}_{spec}\boldsymbol{W}^{V}_2 \right ),
\label{p_specific}
\end{equation}

\begin{equation}
\begin{aligned}
&\operatorname{MH}( \mathbf{Q}, \mathbf{K}, \mathbf{V}  )  = \operatorname{Concat}(\boldsymbol{head}_1,...,\boldsymbol{head}_n)\mathbf{W}^{O}, \\
&{\rm where} \ \boldsymbol{head}_{i} = \operatorname{Attention}( \mathbf{Q W}_{i}^{Q},\mathbf{K W}_{i}^{K}, \mathbf{V W}_{i}^{V}  ),
\label{MH}
\end{aligned}
\end{equation}


\begin{equation}
\operatorname{Attention}(\boldsymbol{Q},\boldsymbol{K},\boldsymbol{V}) = \operatorname{Softmax}\left (  \frac{\mathbf{Q K}^{T} }{\sqrt{d_{k}} } \right )\boldsymbol{V},
\label{Attention}
\end{equation}
where $\boldsymbol{h}_{\mathcal{S}_a^i}$ denotes the sequence representation of $\mathcal{S}_a^i$, $n$ is the head number of the attention network, and $\boldsymbol{W}^Q_{1}$, $\boldsymbol{W}^K_{1}$, $\boldsymbol{W}^V_{1}$, $\boldsymbol{W}^Q_{2}$, $\boldsymbol{W}^K_{2}$, and $\boldsymbol{W}^V_{2}$ are learnable parameters within the network.
$\text{Attention}(\cdot)$ is a Scaled Dot-Product Attention function that computes attention scores by taking the dot product of $\boldsymbol{Q}$ and $\boldsymbol{K}$, then applies the Softmax function to obtain attention weights, and finally uses these weights to weight $\boldsymbol{V}$ to obtain the weighted output.
$\text{Softmax}(\cdot)$ is a normalization function that transforms a set of input values into the form of a probability distribution.

\textbf{Domain Information Fusion.}
To more comprehensively represent domain information, we fuse the sequence representation $\boldsymbol{h}_{\mathcal{S}_a^i}$ with the domain-shared prompts $\boldsymbol{P}_{shared}^{'}$ and domain-specific prompts $\boldsymbol{P}_{spec}^{'}$ in Eq.~(\ref{p_share}) and Eq.~(\ref{p_specific}) by the concatenation operation:
\begin{equation}
\boldsymbol{h}^{'}_{\mathcal{S}_a^{i}} = \boldsymbol{h}_{\mathcal{S}_a^{i}} \oplus \boldsymbol{P}_{shared}^{'} \oplus \boldsymbol{P}_{spec}^{'},
\label{concat}
\end{equation}
where $\boldsymbol{h}^{'}_{\mathcal{S}_a^{i}} \in \mathbb{R}^{3*d_V}$ is the final representation of the given sequence. We feed the sequence representation $\boldsymbol{h}^{'}_{\mathcal{S}_a^{i}}$ into the MLP network to compress the $\boldsymbol{h}^{'}_{\mathcal{S}_a^{i}}$ dimension to $d_V$.
\begin{equation}
\boldsymbol{h}^{''}_{\mathcal{S}_{a}^{i}}=\sigma \left ( \boldsymbol{h}^{'}_{\mathcal{S}_a^{i}} \cdot \boldsymbol{W}_{1} +  \boldsymbol{b}_{1} \right ) \cdot \boldsymbol{W}_{2} + \boldsymbol{b}_{2},
\label{mlp}
\end{equation}
where $\sigma(\cdot)$ is a non-linear activation function and $\boldsymbol{W}_1$,$\boldsymbol{W}_2$,$\boldsymbol{b}_1$,$\boldsymbol{b}_2$ are all learnable parameters. Finally, we obtain the sequence representation $\boldsymbol{h}^{''}_{\mathcal{S}_{a}^{i}} \in \mathbb{R}^{d_V}$ after prompt enhancement.

Similarly, we can obtain the item representation $\boldsymbol{h}^{''}_{v_{b}^{i}}$ by viewing it ($\boldsymbol{h}_{v_{b}^{i}}$) as a sequence of length 1, undergoing Eqs.~(\ref{p_share}), (\ref{p_specific}), (\ref{MH}), (\ref{Attention}), (\ref{concat}), and (\ref{mlp}).

\begin{figure*}[!t]
    \centering
    \includegraphics[width=14cm]{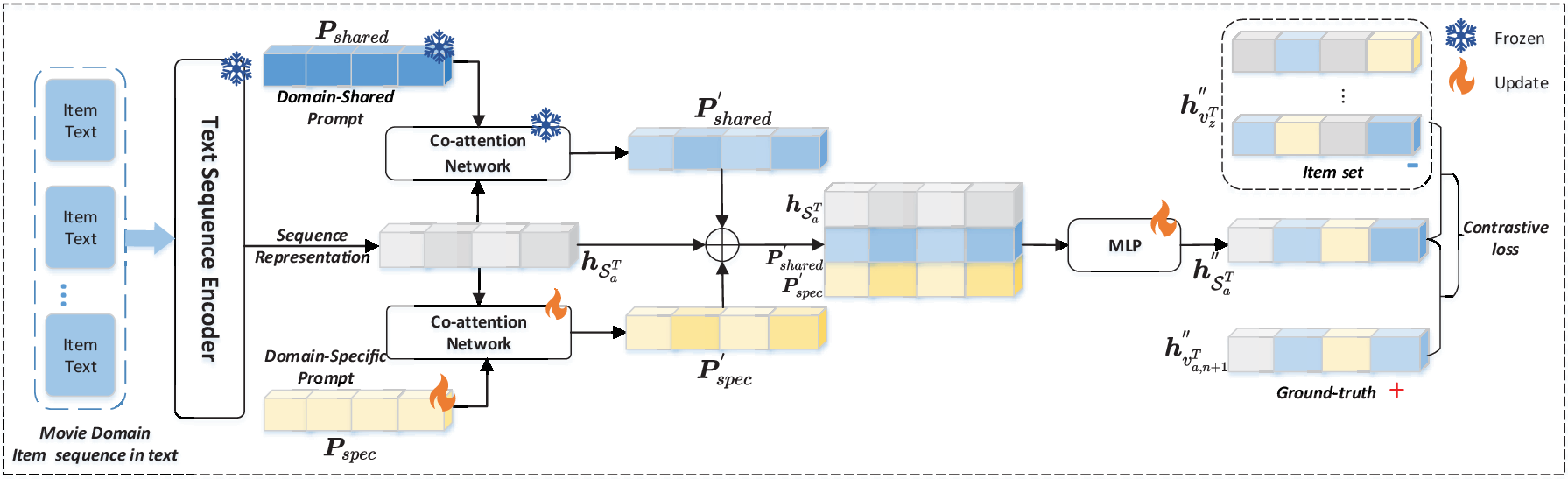}
    \caption{
    The workflow of \ac{TCPLP} in the prompt-tuning stage. During this stage, we freeze the text sequence encoder, domain-shared prompt parameters, and the co-attention network based on domain-shared prompt. We only update the domain-specific prompt, the co-attention network based on domain-specific prompt, and the MLP network.}
    \label{fig:finetune}
\end{figure*}

\begin{algorithm}

\small
    \caption{\fontsize{9.4bp}{17bp}$\proc{The training process of \ac{TCPLP}}.$}
    \label{alg:algorithm1}
    \KwIn{The text of interaction sequences $\mathcal{S}_{a}^{i}$ from all domains, The text of interaction sequence $\mathcal{S}_{a}^{T}$ from the target domain $D_T$}
    \KwOut{Next-item predictions for each user in the target domain;}
    
    \BlankLine
    \textbf{Stage 1: Pre-training with prompts learning}\;
    Initialized the text sequence encoder parameters\;
    Initialized the domain-shared prompts and domain-specific prompts\;

    \For{each epoch}{
        \For{each batch in $\mathcal{S}_{a}^{i}$}{
            Obtain the sequence and item representation $\boldsymbol{h}_{\mathcal{S}_{a}^{i}}$ and $\boldsymbol{h}_{v^{i}_{b}}$ via Eqs.~(\ref{hs}) and (\ref{hv})\;
            Obtain the enhanced domain-shared prompts $\boldsymbol{P}_{shared}^{'}$ in co-attention network via Eq.~(\ref{p_share})\;
            Obtain the enhanced domain-specific prompts $\boldsymbol{P}_{spec}^{'}$ in co-attention network via Eq.~(\ref{p_specific})\;
            Obtain the prompt-enhanced sequence representation $\boldsymbol{h}^{''}_{\mathcal{S}_{a}^{i}}$ through representation fusion via Eqs.~(\ref{concat}) and (\ref{mlp})\;
            Calculate the contrastive loss in the pre-training stage via Eq.~(\ref{loss})\;
            Update model parameters using gradient descent\;
        }
    }
    \BlankLine
    \textbf{Stage 2: Prompt-tuning}\;
        Freeze the text sequence encoder parameters\;
        Freeze the domain-shared prompts\;
    \For{each epoch}{
        Obtain all item representation $\boldsymbol{h}_{v_{b}^{T}}$ of the target domain $D_T$ via Eq.~(\ref{hv})\;
        \For{each batch in $\mathcal{S}_{a}^{T}$}{
            Obtain the sequence representation $\boldsymbol{h}_{\mathcal{S}_{a}^{T}}$ via Eq.~(\ref{hs})\;
            Obtain the enhanced domain-shared prompts $\boldsymbol{P}_{shared}^{'}$ in co-attention network via Eq.~(\ref{p_share})\;
            Obtain the enhanced domain-specific prompts $\boldsymbol{P}_{spec}^{'}$ in co-attention network via Eq.~(\ref{p_specific})\;
            Obtain the prompt-enhanced sequence representation $\boldsymbol{h}^{''}_{\mathcal{S}_{a}^{T}}$ through representation fusion via Eqs.~(\ref{concat}) and (\ref{mlp})\;
            Calculate the contrastive loss in the prompt-tuning stage via Eq.~(\ref{finetune_loss})\;
            Update model parameters using gradient descent\;
        }
    }
    \BlankLine
\end{algorithm}

\subsection{Optimization in the Pre-training Stage}

We adopt a two-stage learning strategy, pre-train \& prompt tuning, for \ac{MNCSR}, where the pre-training stage responses for knowledge pre-learning and prompting are designed to adapt the learned knowledge to the target domain (the training process of our \ac{TCPLP} method is shown in Algorithm \ref{alg:algorithm1}). In the pre-training stage, we encode multi-domain sequence texts using the text sequence encoder, where domain-shared prompts are utilized to capture domain invariant features in all domains.
In the prompt-tuning stage, we only update the domain-specific prompts to meet the specific distribution of the target domain.

\textbf{Optimization Objective.}
During the pre-training stage, we employ a comprehensive optimization strategy that simultaneously optimizes the parameters of the text sequence encoder, domain-shared prompt, domain-specific prompt, co-attention network parameters, and MLP parameters. 
Our pre-training task is to predict the next item in a sequence. We employ a contrastive learning approach to optimize our objective.
Specifically, for each sequence, we consider the ground truth item in the sequence as positive instances, while treating the ground truth item of other sequences within the same batch as the negative instances.
The reasons for using contrastive learning to optimize the objective are as follows:
1) Our method not only focuses on ranking items but also needs to capture domain semantics from different domains. 
Contrastive loss provides us with an effective way to distinguish between positive and negative instances during the learning process, which is crucial for learning discriminative features from multiple domains.
It allows the model to better understand the similarities and differences between items from different domains, thus enhancing the generalization ability of the model.
2) In the pre-training stage, contrastive loss plays an important role in facilitating the learning of prior domain knowledge. Specifically, we consider the ground truth item in a sequence as a positive instance and the ground truth item of other sequences in the batch as a negative instance.
As the negative instances are derived from multiple domains, this can enhance the cross-domain semantic fusion ability of our method and help to improve the quality of the generated representations.
The training loss is formulated as:
\begin{equation}
\mathcal{L}_{pre}= - \operatorname{log}\frac{\operatorname{exp}\left ( \boldsymbol{h}^{''}_{\mathcal{S}_{a}^{i}}\cdot \boldsymbol{h}^{''}_{v_{a,n+1}^{i}}/\tau \right ) }{ {\textstyle \sum_{z=1}^{B}}\operatorname{exp}\left ( \boldsymbol{h}^{''}_{\mathcal{S}_{a}^{i}}\cdot \boldsymbol{h}^{''}_{v_{u_{z},n+1}^{i}}/\tau \right )  },
\label{loss}
\end{equation}
where $B$ is the batch size, $\boldsymbol{h}^{''}_{\mathcal{S}_{a}^{i}}$ is the sequence representation of user sequence $\mathcal{S}_{a}^{i}$ in domain $D_i$ from Eq.~(\ref{mlp}), and 
$\boldsymbol{h}^{''}_{v_{a,n+1}^{i}}$ represents the ground truth item representation of the user sequence $\mathcal{S}_{a}^{i}$, and $\boldsymbol{h}^{''}_{v_{u_{z},n+1}^{i}}$ is the representation of ground truth items of other sequences within the batch.

\subsection{The Prompt-tuning Stage}
After the pre-training stage, we obtain the domain-shared prompts and a text sequence encoder trained in all domains. As these two components are respectively designed to comprehend recommendation text sequences and capture invariant knowledge from multiple domains, we tend to freeze them in the fine-tuning stage.
The target of the prompt-tuning stage is to transfer invariant knowledge across domains, with the aim of alleviating the sparsity issue in the target domain.
Specifically, during this stage, we freeze the parameters in the text sequence encoder, domain-shared prompts, and the co-attention network for domain-shared prompt encoding. We only fine-tune the parameters in the domain-specific prompts and the corresponding prompt encoder to adapt the pre-learned knowledge to meet the specific data distribution of the target domain (as shown in Fig.~\ref{fig:finetune}).

In the prompt-tuning stage, the same contrastive learning optimization strategy is used as in the pre-training stage. The difference is that we select the negative instances from all the items in the target domain. The loss function of this stage is defined as:

\begin{equation}
\mathcal{L}_{pt}= -\operatorname{log}\frac{\operatorname{exp}\left ( \boldsymbol{h}^{''}_{\mathcal{S}_{a}^{T}}\cdot \boldsymbol{h}^{''}_{v_{a,n+1}^{T}}/\tau \right ) }{ {\textstyle \sum_{z=1}^{m}}\operatorname{exp}\left ( \boldsymbol{h}^{''}_{\mathcal{S}_{a}^{T}}\cdot \boldsymbol{h}^{''}_{v_{z}^{T}}/\tau \right )  },
\label{finetune_loss}
\end{equation}
where $m$ is the number of items in the target domain, $\boldsymbol{h}^{''}_{\mathcal{S}_{a}^{T}}$ is the sequence representation of the user sequence $\mathcal{S}_{a}^{T}$ in the target domain $D_T$, $\boldsymbol{h}^{''}_{v_{a,n+1}^{T}}$ is the ground truth item representation of the target domain sequence $\mathcal{S}_{a}^{T}$, and $\boldsymbol{h}^{''}_{v_{z}^{T}}$ is the item representation of all the items in the target domain.

\section{Experimental Setup}
In this section, we first introduce the research questions to be answered in the experiments. 
Then, we provide the details of our utilized datasets, the evaluation method, baselines, and the implementation of our approach.

\subsection{Research Questions}
We comprehensively evaluate our \ac{TCPLP} approach by answering the following research questions:

\begin{itemize}
    \item[\textbf{RQ1}] How does \ac{TCPLP} perform compared to the state-of-the-art baselines?
    \item[\textbf{RQ2}] How do the key components of \ac{TCPLP}, i.e., the domain-shared and domain-specific prompts, co-attention network, pre-training and prompt-tuning strategies, contribute to the performance of \ac{TCPLP}?
    \item[\textbf{RQ3}] What is the impact of different pre-trained language models on \ac{TCPLP}?
    \item[\textbf{RQ4}] What is the effect of using different optimization losses on \ac{TCPLP}?
    \item[\textbf{RQ5}] What are the effects of key hyper-parameters on \ac{TCPLP}?

\end{itemize}

\subsection{Datasets and Evaluation Metrics}
We conducted experimental evaluations on two subsets of the Amazon Review dataset~\cite{amazon} (\textit{Office-Arts} and \textit{Cell-Toys-Automotive}) and a set of cross-platform datasets (\textit{OR-Pantry-Instruments}).
The cross-platform dataset is composed of the Online Retail (\textit{OR}) dataset\footnote{https://www.kaggle.com/carrie1/ecommerce-data} and Pantry and Instruments domains from the Amazon Review dataset~\cite{amazon}. 
The Amazon Review dataset~\cite{amazon} contains user review information and product metadata for products in different domains, including userid, itemid, item title, and review time, etc.
The Online Retail dataset is a transnational data set that contains all the transactions occurring between 01/12/2010 and 09/12/2011 for a UK-based and registered non-store online retail. 
The sub-collection \textit{Office-Arts} only references users' behaviors in \textit{Office} and \textit{Arts}.
This is to evaluate our recommendation ability in two domains.
The sub-collection \textit{Cell-Toys-Automotive} utilizes behavioral knowledge from three domains of \textit{Cell}, \textit{Toys}, and \textit{Automotive}. This is to evaluate the enhancement of recommendation capabilities in the target domain by leveraging multi-domain knowledge.
The cross-platform dataset (\textit{OR-Pantry-Instruments}) utilizes multi-platform behavioral knowledge from Online Retail and from \textit{Pantry} and \textit{Instruments} in Amazon. This is to evaluate the enhancement of recommendation capabilities in the target domain using cross-platform behavioral knowledge.

Since our target is for sequential recommendation, we sort items by their chronological order in each user to get her interaction sequence. 
To satisfy the non-overlapping setting, we only keep the users that are disjoint among different domains for training and testing.
To filter out invalid information, we further remove the users whose interaction sequence length is less than 5 and items interacted with by fewer than 5 users.
We take the last item of each sequence as the ground truth.
We utilize the tile information of each item as its content representation.
In the pre-training stage, we use the training data of all domains for training. In the prompt-tuning stage, we only use the training data of the target domain for prompt-tuning. The statistics of our dataset are presented in Table~\ref{tab:dataset_statistics}.

We employ the widely used Recall@K and NDCG@K as our evaluation metrics. Here, K represents the number of recommended items returned to the user with the highest probability. 
Specifically, we set K to 10 and 20 in our experiments.

\begin{itemize}       
    \item Recall@K: This evaluation metric measures the proportion of relevant items successfully recommended by the model among the top-K recommended items. It is more capable of assessing the comprehensiveness of the recommendation system.
    \item NDCG@K: This metric measures the quality of the ranking of relevant items among the top-K recommended items. It not only focuses on the number of relevant items in the recommendations but also considers their positions. Therefore, it can more precisely evaluate the performance of the recommendation system.
\end{itemize}

\begin{table}
    \centering
    \footnotesize
     \caption{The statistics for the \textit{Office-Arts}, \textit{Cell-Toys-Automotive}, and \textit{OR-Pantry-Instruments} datasets.}
    \scalebox{1.2}{
    \setlength{\tabcolsep}{1.5mm}{
        \begin{tabular}{l|c|c|c|c|c|c|c|c}
        \toprule
        \multicolumn{1}{c|}{\multirow{2}[1]{*}{\textbf{Details}}}&\multicolumn{2}{c|}{\textit{Office-Arts}}&       \multicolumn{3}{c|}{\textit{Cell-Toys-Automotive}}&       \multicolumn{3}{c}{\textit{OR-Pantry-Instruments}} \\
        \cmidrule{2-9}
        & \textit{Office} & \textit{Arts}
        & \textit{Cell} & \textit{Toys} & \textit{Automotive} 
        & \textit{OR} & \textit{Pantry} & \textit{Instruments}\\
        \midrule
        \#Users &87,346 &45,486 &88,348 & 112,307 & 93,052 & 16,520 & 13,101 & 24,962\\
        \#Items &25,986 &21,019 &31,028 & 49,575 & 43,622 & 3,460 & 4,898 & 9,964\\
        \#Inters &684,837&395,150&621,498&960,501&774,313 & 385,529 & 124,956 & 207,568\\
        \#Avg.n &7.84&8.68&7.03&8.54&8.31 & 23.33&9.53&8.31\\
        \#Train &76,991&35,551&70,678&89,845&74,441&13,216&10,480&19,969\\
        \#Val &778&358&8,834&11,230&9,305&1,652&1,310&2,496\\
        \#Test &9,577&9,577&8,836&11,232&9,306&1,652&1,311&2,497\\
        \bottomrule
        \end{tabular}
    }}
    \label{tab:dataset_statistics}
\end{table}

\subsection{Baselines}
We compare \ac{TCPLP} with four types of baselines, i.e., the ID-only sequential recommendation methods, the text-only sequential recommendation methods, the ID-Text sequential recommendation methods, and the cross-domain recommendation methods.

1) ID-only sequential recommendation:
\begin{itemize}
 \item \textbf{POP}: 
This is a simple and effective baseline that recommends top-$K$ items based on their popularity.
 \item \textbf{GRU4Rec}~\cite{GRU4Rec}: This RNN-based session recommendation method addresses the limitations of matrix factorization methods in domains with short session data. By modeling entire sessions, this approach offers more accurate recommendations.
 \item \textbf{SASRec}~\cite{SASRec}: This is a self-attention-based sequence recommendation model that uses a self-attention network to capture user preferences. It is a classical method for single-domain sequential recommendation.
\end{itemize}

2) Text-only sequential recommendation:
\begin{itemize}
 \item \textbf{UniSRec}~\cite{text_UniSRec}: This is a universal sequence representation learning approach for sequential recommenders, using item descriptions to create transferable representations across recommendation scenarios. Through lightweight item encoding and contrastive pre-training tasks, UniSRec enables effective transfer to new recommendation domains.
 \item \textbf{VQ-Rec}~\cite{text_VQRec}: This is a novel approach for transferable sequential recommenders, aiming to address issues with existing methods that tightly bind item text and representations. VQ-Rec uses Vector-Quantized item representations, mapping item text to discrete indices, and then using these indices to derive item representations, improving transferability.
 \item \textbf{Recfomer}~\cite{text_recformer}: This is a novel framework for sequential recommendation, which leverages language representations to model user preferences and item features.

\end{itemize}

3) Cross-domain recommendation:
\begin{itemize}
 \item \textbf{$\pi$-net}~\cite{PINET}: This method solves the item prediction task given the mixed multi-user behaviors of shared accounts, and transfers cross-domain information through a parallel information sharing network. This \ac{CDR} method requires all the accounts to be fully overlapped between domains. This is different from our MNCSR task, which does not require any overlapping entities.
 \item \textbf{PLCR}~\cite{soft_prompt_PLCR}: This is a dual-target \ac{CDR} based on the automatic prompt engineering paradigm, and aims to address the non-overlapping challenge in the \ac{NCSR} task. Although this method also solves the non-overlapping recommendation task, it only focuses on two domains (a source domain and a target domain), and it is difficult to directly extend it to multi-source domain tasks.
 \item \textbf{PFCR}~\cite{PFCR}: This is a federated \ac{CDR} framework that establishes item-level semantic alignment through shared vector-quantized codebooks, enabling privacy-preserving knowledge transfer by progressively optimizing discrete item representations across domains. 
 \item \textbf{MCRPL}~\cite{soft_prompt_MCRPL}: This method develops an ID-based prompt learning paradigm for solving \ac{MNCSR}.
\end{itemize}

4) ID-Text sequential recommendation:
\begin{itemize}
 \item \textbf{S$^3$Rec}~\cite{baseline_S3Rec}: This is a self-supervised learning approach for sequential recommendation with a self-attentive neural architecture. It enhances data representations by utilizing intrinsic data correlations in a pre-training paradigm.
\end{itemize}

\subsection{Implementation Details}
We implement \ac{TCPLP} based on PyTorch and accelerate the training process using a GeForce GTX TitanX GPU. The language model, Longformer\footnote{https://huggingface.co/allenai/longformer-base-4096} and Bert\footnote{https://huggingface.co/google-bert/bert-base-uncased}, are from the Huggingface platform. We set the maximum number of items in the interaction sequence to 50. For Longformer, we set the maximum token number in the sequence text to 1024. For BERT, since its maximum number of tokens is 512, we set it to 512. The hyper-parameter $d_W$ is set to 2 for both domain-shared and domain-specific prompts. We set the dimension of the sequence representation and prompt embedding to 768 for computation convenience. Our approach consists of two parts: pre-training and prompt-tuning. During the pre-training stage, we set the batch size to 12. In the prompt-tuning stage, the batch size is set to 16. We use Adam as our optimizer for optimization, with the learning set to 5e-5 and the temperature coefficient set to 0.05.

To ensure a fair comparison with the baselines, we conduct a grid search on the key hyper-parameters of all the baselines. 
The hyper-parameter settings of these regarding the implementation details of the baselines, we describe them as follows:

1) For ID-only sequential recommendation (i.e., GRU4Rec, SASRec, POP), since these methods are single-domain sequential recommendation methods, we train and test them using only the target domain sequence.
For GRU4Rec, we perform a grid search for key hyper-parameters in dropout rate and learning rate over all our datasets. We tune the dropout rate in \{0.1, 0.2, 0.3, 0.4, 0.5\} and the learning rate in \{0.01, 0.05, 0.1, 0.15, 0.2\}. For SASRec, we perform a grid search for latent dimensionality in the key hyper-parameters. We tune the latent dimensionality in \{64, 128, 256, 512\}.
2) For text-only sequential recommendation methods (i.e., UniSRec, VQ-Rec, Recformer), we utilize item titles to represent items. Since these three methods use a pre-training and fine-tuning strategy, we adopt the same strategy as TCPLP: training using all domain sequence data in the pre-training phase and fine-tuning the target domains in the fine-tuning phase. 
For UniSRec, we perform a grid search for embedding dimension and learning rate in the key hyper-parameters. We tune the embedding dimension in \{64, 128, 256, 512\} and the learning rate in \{0.0003, 0.001, 0.003, 0.01\}.
For VQ-Rec, we perform a grid search for learning rate and permutation learning epochs in the key hyper-parameters. We tune the learning rate in \{0.0003, 0.001, 0.003, 0.01\} and the permutation learning epochs in \{3, 5, 10\}.
For RecFormer, We tune $\lambda$ in \{0.01, 0.05, 0.1, 0.2, 0.5\}.
3) For cross-domain recommendation methods (i.e., $\pi$-Net, PLCR, MCRPL, PFCR), we conduct experiments on CSR methods based on overlapping entities and CSR based on non-overlapping entities respectively.
For $\pi$-Net, since this work is a CSR method based on overlapping entities, we randomly pair user interaction sequences in different domains as shared user sequences. For NCSR methods (PLCR, MCRPL), we utilize item IDs as item representations. 
For $\pi$-Net, we perform a grid search for K (the number of latent users) and the learning rate in the key hyper-parameters. We tune K in \{1, 2, 3, 4, 5\} and the learning rate in \{0.0001, 0.001, 0.01\}.
For PLCR, we perform grid search for the key hyper-parameters dropout ratio and domain-invariant contexts. We tune the dropout ratio in \{0.1, 0.2, 0.3, 0.4, 0.5\} and the domain-invariant context in the range of [1, 8]. 
To meet our \ac{MNCSR} setting, we integrate all the training data from the source domains to achieve multi-source domain enhancement.
For MCRPL, we perform a grid search for the key hyper-parameters $\lambda$ and $L_p$. We tune $\lambda$ in the range of [0.01, 0.1] and $L_p$ in \{1, 2, 3, 4, 5\}.
For PFCR, we perform a grid search for the learning rate and the head of the sequence encoder. We tune the learning rate in \{0.0001, 0.001, 0.01\} and the head number in the range of [1, 5].
4) For ID-Text sequential recommendation methods (i.e., S$^{3}$Rec), we utilize both the textual data of the items and the IDs of the items.
For S$^{3}$Rec, we perform a grid search for the key hyper-parameter mask ratio and the weights of AAP, MIP, MAP, and SP.
We tune L in \{0.1, 0.2, 0.5, 0.8, 1.0\} and the weights of AAP, MIP, MAP and SP in the range [0, 1].

\section{Experimental Results (RQ1)}

\begin{table*}
\centering
\small
\caption{The overall performance of the baseline methods and the proposed framework is evaluated on the \textit{Office-Arts}, \textit{Cell-Toys-Automotive} and \textit{OR-Pantry-Instruments} dataset. The best performance is highlighted in bold, and the second-best result is underlined. \ac{TCPLP} (L) stands for text encoder using Longformer for encoding. \ac{TCPLP} (B) stands for text encoder using BERT for encoding.}
\vspace{3mm}
\label{tab:result}
\renewcommand{\arraystretch}{1.3}
\scalebox{0.63}{
\setlength{\tabcolsep}{1.5mm}{
\begin{tabular}{llccccccccccccccc}
\toprule
\multicolumn{1}{c}{\textbf{}} &    \multicolumn{1}{c}{\textbf{}}   & \multicolumn{3}{c}{\textbf{ID-Only}}  &\multicolumn{3}{c}{\textbf{Text-Only}}  & \multicolumn{4}{c}{\textbf{Cross-domain}} & \multicolumn{1}{c}{\textbf{ID-Text}} & \multicolumn{2}{c}{\textbf{Ours}}  \\ \cmidrule(lr){3-5} \cmidrule(lr){6-8} \cmidrule(lr){9-12} \cmidrule(lr){13-13} \cmidrule(lr){14-15} 
\multicolumn{1}{c}{\textbf{Dataset}} & \multicolumn{1}{c}{\textbf{Metric}} & \textbf{GRU4Rec} & \textbf{SASRec} & \textbf{POP} & \textbf{UniSRec} & \textbf{VQ-Rec}  & \textbf{Recformer}        & \textbf{$\pi$-net}       & \textbf{PLCR} & \textbf{MCRPL} & \textbf{PFCR} & \textbf{S$^3$Rec} & \textbf{\ac{TCPLP}(L)} & \textbf{\ac{TCPLP}(B)}                                     \\ \midrule
\multirow{4}{*}{\textit{Office}} & Recall@10 & 0.0542 & 0.0653 & 0.0257 & 0.0682 & 0.0733  & 0.0757  & 0.0608 & 0.0675 & 0.0673 & 0.0707 & 0.0624  & \textbf{0.0913*} & \underline{0.0894}    \\
& NDCG@10 & 0.0422 & 0.0464 & 0.0259 & 0.0468 & 0.0510  & 0.0473 & 0.0421  & 0.0477 & 0.0489 & 0.0504 & 0.0414 & \textbf{0.0593*}  & \underline{0.0587} \\
& Recall@20 & 0.0621  & 0.0754  & 0.0379 & 0.0843 & 0.0899  & 0.0928 & 0.0745 & 0.0763 & 0.0785 & 0.0885 & 0.0711 & \textbf{0.1159*}  & \underline{0.1128}  \\ 
& NDCG@20 & 0.0446  & 0.0487  & 0.0362 & 0.0509 & 0.0550  & 0.0511 & 0.0471 & 0.0488 & 0.0518 & 0.0537 & 0.0442 & \textbf{0.0655*}  & \underline{0.0646} \\ \cline{2-15}

\multirow{4}{*}{\textit{Arts}} & Recall@10 & 0.0292 & 0.0530 & 0.0178 & 0.0621 & 0.0577  & 0.0683 & 0.0451 & 0.0423 & 0.0485 & 0.0613 & 0.0494 & \textbf{0.0751*}  & \underline{0.0720}   \\
& NDCG@10 & 0.0193 & 0.0364 & 0.0115 & 0.0372 & 0.0376  & 0.0422 & 0.0317 & 0.0311 & 0.0335 & 0.0370 & 0.0353 & \textbf{0.0451*}  & \underline{0.0438}   \\
& Recall@20 & 0.0368  & 0.0613  & 0.0221 & 0.0747 & 0.0688 & 0.0873 & 0.0563 & 0.0536 & 0.0516 & 0.0724 & 0.0587  & \textbf{0.1007*}  & \underline{0.0987}    \\ 
& NDCG@20 & 0.0211  & 0.0382 & 0.0142 & 0.0414 & 0.0411  & 0.0474 & 0.0341 & 0.0343 & 0.0350 & 0.0408 & 0.0385 & \textbf{0.0516*}  & \underline{0.0505}  \\ \midrule

\multirow{4}{*}{\textit{Cell}}& Recall@10 & 0.0321 & 0.0425 & 0.0247 & 0.0593 & 0.0536  & 0.0634 & 0.0457 & 0.0499 & 0.0385 & 0.0589 & 0.0518 & \textbf{0.0791*} & \underline{0.0746}   \\
& NDCG@10 & 0.0224 & 0.0269 & 0.0167 & 0.0346 & 0.0290 & 0.0393 & 0.0267 & 0.0278 & 0.0263 & 0.0339 & 0.0289 & \textbf{0.0457*} & \underline{0.0444}\\
& Recall@20 & 0.0405  & 0.0517  & 0.0298 & 0.0850 & 0.0721  & 0.0923 & 0.0558 & 0.0556 & 0.0458 & 0.0837 & 0.0628 & \textbf{0.1148*} & \underline{0.1056}   \\ 
& NDCG@20 & 0.0252  & 0.0287  & 0.0204 & 0.0400 & 0.0341  & 0.0462 & 0.0290 & 0.0309 & 0.0291 & 0.0395 & 0.0301 &\textbf{0.0547*}  & \underline{0.0522}   \\ \cline{2-15}

\multirow{4}{*}{\textit{Toys}}& Recall@10 & 0.0172 & 0.0312 & 0.0156 & 0.0557 & 0.0455 & \textbf{0.0703} & 0.0350 & 0.0393 & 0.0317 & 0.0533 & 0.0390 & \underline{0.0693} & 0.0680   \\
& NDCG@10 & 0.0109 & 0.0181 & 0.0121 & 0.0295 & 0.0238  & \underline{0.0369} & 0.0187 & 0.0197 & 0.0189 & 0.0290 & 0.0194 & \textbf{0.0373*} & 0.0355 \\
& Recall@20 & 0.0221  & 0.0366  & 0.0201 & 0.0782 & 0.0621  & \underline{0.0990} & 0.0337 & 0.0392 & 0.0374 & 0.0745 & 0.0388 & \textbf{0.0994*}  & 0.0960 \\ 
& NDCG@20 & 0.0122  & 0.0195  & 0.0132 & 0.0354 & 0.0297  & \underline{0.0443} & 0.0195 & 0.0220 & 0.0202 & 0.0347 & 0.0281 &\textbf{0.0448*}  & 0.0425  \\ \cline{2-15}

\multirow{4}{*}{\textit{Automotive}} & Recall@10 & 0.0221 & 0.0336 & 0.0173 & 0.0452  & 0.0434 & 0.0472 & 0.0328 & 0.0362 & 0.0341
 & 0.0446 & 0.0389 & \underline{0.0519}  & \textbf{0.0548*}  \\
& NDCG@10 & 0.0159 & 0.0211 & 0.0151 & 0.0262 & 0.0247  & 0.0281 & 0.0215 & 0.0225 & 0.0209 & 0.0258 & 0.0218 &\underline{0.0297}  & \textbf{0.0320*}\\
& Recall@20 & 0.0277  & 0.0431  & 0.0247 & 0.0587 & 0.0542 & 0.0635 & 0.0420 & 0.0435 & 0.0443 & 0.0571 & 0.0489 & \underline{0.0716}  & \textbf{0.0739*} \\ 
& NDCG@20 & 0.0177  & 0.0227 & 0.0170 & 0.0291 & 0.0276  & 0.0318 & 0.0225 & 0.0235 & 0.0230 & 0.0286 & 0.0258 & \underline{0.0347} & \textbf{0.0368*}  \\ \midrule

\multirow{4}{*}{\textit{OR}}& Recall@10 & 0.0309 & 0.0696 & 0.0125 & 0.0975 & 0.0884  & 0.1404 & 0.0606 & 0.0608 & 0.0703 & 0.0966 & 0.0690 & \textbf{0.1547*} & \underline{0.1432}   \\
& NDCG@10 & 0.0148 & 0.0286 & 0.0087 & 0.0463 & 0.0397  & 0.0667 & 0.0265 & 0.0292 & 0.0356 & 0.0454 & 0.0288 & \textbf{0.0734*} & \underline{0.0680}\\
& Recall@20 & 0.0551  & 0.1150  & 0.0253 & 0.1610 & 0.1447  & 0.2117 & 0.1090 & 0.1087 & 0.1110 & 0.1662 & 0.1053 & \textbf{0.2378*} & \underline{0.2166}   \\ 
& NDCG@20 & 0.0210  & 0.0399  & 0.0098 & 0.0621 & 0.0539  & 0.0853 & 0.0342 & 0.0406 & 0.0465 & 0.0634 & 0.0379 &\textbf{0.0945*}  & \underline{0.0866}   \\ \cline{2-15}

\multirow{4}{*}{\textit{Pantry}}& Recall@10 & 0.0290 & 0.0267 & 0.0112 & 0.0366 & 0.0206 & 0.0416 & 0.0215 & 0.0287 & 0.0274 & 0.0348 & 0.0282 & \textbf{0.0493*} & \underline{0.0447}   \\
& NDCG@10 & 0.0130 & 0.0125 & 0.0058 & 0.0173 & 0.0101  & \underline{0.0202} & 0.0118 & 0.0133 & 0.0126 & 0.0167 & 0.0140 & \textbf{0.0254*} & 0.0196 \\
& Recall@20 & 0.0496  & 0.0450  & 0.0271 & 0.0595 & 0.0374  & 0.0709 & 0.0429 & 0.0416 & 0.0373 & 0.0581 & 0.0496 & \underline{0.0709*}  & \textbf{0.0733} \\ 
& NDCG@20 & 0.0181  & 0.0170  & 0.0075 & 0.0231 & 0.0144  & \underline{0.0276} & 0.0165 & 0.0176 & 0.0158 & 0.0223 & 0.0194 &\textbf{0.0309*}  & 0.0268  \\ \cline{2-15}

\multirow{4}{*}{\textit{Instruments}} & Recall@10 & 0.0589 & 0.0837 & 0.0285 & 0.0952  & 0.0781 & 0.0889 & 0.0844 & 0.0811 & 
0.0855 & 0.0937 & 0.0673 & \underline{0.0961}  & \textbf{0.1029*}  \\
& NDCG@10 & 0.0308 & 0.0494 & 0.0131 & 0.0634 & 0.0564  & 0.0510 & 0.0502 & 0.0505 & \underline{0.0644} & 0.0631 & 0.0406 &  0.0612  & \textbf{0.0651*}\\
& Recall@20 & 0.0797  & 0.1065  & 0.0358 & 0.1237 & 0.1061 & 0.1133 & 0.1102 & 0.1072 & 0.1033 & 0.1198 & 0.0885 & \underline{0.1239}  & \textbf{0.1298*} \\ 
& NDCG@20 & 0.0360  & 0.0551  & 0.0174 & 0.0692 & 0.0635  & 0.0572 & 0.0562 & 0.0568 & 0.0692 & 0.0679 & 0.0460 & \underline{0.0699} & \textbf{0.0718*} \\ \bottomrule
\end{tabular}
}
}
\begin{tablenotes}  
        \footnotesize       
        \item 
        Significant improvements over baselines are marked with * (Holm-Bonferroni correction, $p<$.05).
\end{tablenotes}
\end{table*}

The experimental results of \ac{TCPLP} and the baseline methods are shown in Table~\ref{tab:result}.
From the experimental results, we can conclude that:
1) Our \ac{TCPLP} method can achieve the best performance on all datasets in almost all metrics, demonstrating the effectiveness of our pre-training \& prompt tuning paradigm in addressing the \ac{MNCSR} task.
2) Compared to text-only methods (i.e., UniSRec, VQ-Rec), our \ac{TCPLP} method can have better recommendation results, showing the advantage of our two-stage learning framework, and the benefit of adapting the pre-learning domain knowledge to the target domain with the help of the co-attention-based prompt learning technique.
3) The performance gap between \ac{TCPLP} and cross-domain methods in both overlapping and non-overlapping scenarios (i.e., $\pi$-net, PLCR, MCRPL, PFCR), demonstrates the availability of text-based item modeling methods, and the effectiveness of co-attention prompt learning paradigms in transferring knowledge from multi-domain to the target domain.
4) The performance of ID-Text-based methods (i.e., S$^{3}$Rec) and text-based methods (i.e., UniSRec, VQ-Rec, RecFormer) is better than that of ID-only methods (i.e., GRU4Rec, SASRec, POP), indicating the usefulness of textual information in enhancing items' representation, which can better capture semantic associations between items and improve the accuracy of recommendations.
\section{Experimental Analysis}
In this section, we first conduct ablation studies to investigate the contribution of different modules to \ac{TCPLP} to answer RQ2. We then examine the influence of various pre-trained language models on the model to answer RQ3. 
We study the influence of different losses on TCPLP to answer RQ4.
Finally, we investigate the influence of key hyper-parameters on \ac{TCPLP} to answer RQ5.

\subsection{Ablation Study (RQ2)}

To show the contribution of different modules to \ac{TCPLP}, we answer RQ2 by comparing \ac{TCPLP} with its six variants.
\begin{itemize}
 \item \textbf{TCPLP-PR}: 
 This is a variant of \ac{TCPLP} that does not go through the pre-training stage and directly performs the prompt-learning on the target domain. This is to evaluate the importance of the pre-training stage in knowledge transfer.
 \item \textbf{TCPLP-PT}: 
 This is a variant of \ac{TCPLP} that does not perform prompt-tuning and only exploits the pre-trained knowledge to directly evaluate the target domain. It is used to evaluate the effectiveness of the prompt-tuning stage.
 \item \textbf{TCPLP-CA}: 
 This is another variant of \ac{TCPLP} without the co-attention network module. This is to evaluate the effectiveness of the co-attention network.
 \item \textbf{TCPLP-SH}: This is another variant of \ac{TCPLP}, where training and testing are conducted without domain-shared prompts. This is used to evaluate whether domain-shared prompts can capture shared features across multiple domains.

 \item \textbf{TCPLP-SP}: This variant of \ac{TCPLP} excludes domain-specific prompts during both training and testing. It is used to evaluate whether domain-specific prompts can adapt to the specific data distribution of the target domain.
 \item \textbf{TCPLP-SSP}: 
This is another variant of \ac{TCPLP}, trained and tested without domain-shared prompts and domain-specific prompts. It serves to evaluate the effectiveness of the prompt module.
\end{itemize}

\begin{table*}
\centering
\small
\caption{Ablation studies on the \textit{Office-Arts} dataset. The best results on it are highlighted with bold text and the second-best result is highlighted with an underline.}
\label{tab:Ablation1}
\renewcommand{\arraystretch}{1.3}
\scalebox{0.9}{
\setlength{\tabcolsep}{1.5mm}{
\begin{tabular}{llccccccc}
\toprule

\multicolumn{1}{c}{\textbf{Dataset}} & \multicolumn{1}{c}{\textbf{Metric}} & \textbf{TCPLP-PR} & \textbf{TCPLP-PT} & \textbf{TCPLP-CA} & \textbf{TCPLP-SH} & \textbf{TCPLP-SP}  & \textbf{TCPLP-SSP}    & \textbf{TCPLP}        \\ \midrule
\multirow{4}{*}{\textit{Office}} & Recall@10 & 0.0696 & 0.0583   & \underline{0.0908} & 0.0877 & 0.0869  & 0.0790 & \textbf{0.0913}    \\
& NDCG@10 & 0.0479 & 0.0372 & \underline{0.0590} & 0.0573 & 0.0581 & 0.0542  & \textbf{0.0593} \\
& Recall@20 & 0.0797  & 0.0745  & \underline{0.1135} & 0.1130 & 0.1104   & 0.1016  & \textbf{0.1159}  \\ 
& NDCG@20 & 0.0504  & 0.0412  & \underline{0.0646}  & 0.0636  & 0.0640 & 0.0599  & \textbf{0.0655} \\ \cline{2-9}

\multirow{4}{*}{\textit{Arts}} & Recall@10 & 0.0620  & 0.0493 & 0.0704 & \underline{0.0730} & 0.0707 & 0.0703  & \textbf{0.0751}   \\
& NDCG@10 & 0.0370 & 0.0285 & 0.0423  & 0.0426 & 0.0432 & \underline{0.0445}  & \textbf{0.0451}   \\
& Recall@20 & 0.0773  & 0.0690  & 0.0962 & \underline{0.0996} & 0.0962 & 0.0922 & \textbf{0.1007}    \\ 
& NDCG@20 & 0.0409  & 0.0335  & 0.0487 & 0.0493 & 0.0496  & \underline{0.0509}  & \textbf{0.0516} \\ \bottomrule
\end{tabular}
}
}

\end{table*}

\begin{table*}
\centering
\small
\caption{Ablation studies on the \textit{Cell-Toys-Automotive} dataset. The best results on it are highlighted with bold text and the second-best result is highlighted with an underline.}
\label{tab:Ablation2}
\renewcommand{\arraystretch}{1.3}
\scalebox{0.87}{
\setlength{\tabcolsep}{1.5mm}{
\begin{tabular}{llccccccc}
\toprule

\multicolumn{1}{c}{\textbf{Dataset}} & \multicolumn{1}{c}{\textbf{Metric}} & \textbf{TCPLP-PR} & \textbf{TCPLP-PT} & \textbf{TCPLP-CA} & \textbf{TCPLP-SH} & \textbf{TCPLP-SP}  & \textbf{TCPLP-SSP}    & \textbf{TCPLP}        \\ \midrule
\multirow{4}{*}{\textit{Cell}} & Recall@10 & 0.0458 & 0.0449  & \underline{0.0756} & 0.0734 & 0.0716  & 0.0641 & \textbf{0.0791}    \\
& NDCG@10 & 0.0291 & 0.0248 & \underline{0.0431} & 0.0427 & 0.0420 & 0.0371 & \textbf{0.0457} \\
& Recall@20 & 0.0619  & 0.0671  & \underline{0.1064} & 0.1037 & 0.1021  & 0.0939  & \textbf{0.1148}  \\ 
& NDCG@20 & 0.0332  & 0.0303  & 0.0508   & \underline{0.0504}  & 0.0496 & 0.0446  & \textbf{0.0547} \\ \cline{2-9}

\multirow{4}{*}{\textit{Toys}} & Recall@10 & 0.0539  & 0.0462 & \underline{0.0690} & 0.0604 & 0.0640 & 0.0680  & \textbf{0.0693}   \\
& NDCG@10 & 0.0290 & 0.0238 & \underline{0.0365}  & 0.0315 & 0.0332 & 0.0364  & \textbf{0.0373}   \\
& Recall@20 & 0.0748  & 0.0731  & \underline{0.0992} & 0.0892 & 0.0911  & 0.0962  & \textbf{0.0994}  \\ 
& NDCG@20 & 0.0343  & 0.0306  & \underline{0.0442} & 0.0388 & 0.0401  & 0.0435  & \textbf{0.0448} \\ \cline{2-9}

\multirow{4}{*}{\textit{Automotive}} & Recall@10 & 0.0267 & 0.0315 & \underline{0.0516} & 0.0504 & 0.0509 & 0.0495 & \textbf{0.0519}   \\
& NDCG@10 & 0.0179 & 0.0188 & 0.0295  & 0.0295 & \underline{0.0297} & 0.0280 & \textbf{0.0297}   \\
& Recall@20 & 0.0348  & 0.0438  & \underline{0.0711} & 0.0665 & 0.0684  & 0.0658  & \textbf{0.0716}    \\ 
& NDCG@20 & 0.0199  & 0.0219  & \textbf{0.0349} & 0.0335 & 0.0341 & 0.0322  & \underline{0.0347} \\ \bottomrule

\end{tabular}
}
}

\end{table*}

We conduct the ablation studies on both datasets, and report the experimental results in Tables~\ref{tab:Ablation1} and \ref{tab:Ablation2}, from which we can observe that:
1) \ac{TCPLP} outperforms the TCPLP-PR variant on both datasets, indicating the necessity of the pre-training stage and the ability to transfer shared information from multiple domains to enhance the target domain performance. 
2) \ac{TCPLP} has better results than TCPLP-PT, indicating that the effectiveness of the prompt-tuning stage in improving the target domain.
3) \ac{TCPLP} performs better than the TCPLP-CA variant, proving the effectiveness of the co-attention network, which can model the prompts well.
4) \ac{TCPLP} has better performance than the TCPLP-SH variant, indicating that domain-shared prompts can effectively extract shared information across multiple domains as prior knowledge.
5) \ac{TCPLP} has better performance than the TCPLP-SP variant, indicating that tuning domain-specific prompts can effectively adapt to the specific data distribution of the target domain.
6) \ac{TCPLP} has better performance than the TCPLP-SSP variant, showing the effectiveness of domain-shared prompts and domain-specific prompts in extracting domain-invariant features and domain-specific features respectively.

\subsection{Compatibility with Pre-trained Language Models (RQ3)}

We investigate the impact of different pre-trained languages in the text encoder to \ac{TCPLP}, and show the experimental results in Table~\ref{tab:result}, where L stands for that we use Longformer to model the semantic meaning of text sequences, while B represents that we exploit BERT for content learning.
From Table~\ref{tab:result}, we can draw the following conclusion.
Firstly, \ac{TCPLP} is compatible with different pre-trained language models, that is, we can take different language models as a plugin to enhance the ability of our text encoding.
Secondly, the performance of exploiting Longformer for text learning exceeds BERT in most cases.
This is primarily because our item text sequences are usually longer than normal language sentences, and Longformer is more suitable to handle longer sentences than BERT.
We can exploit different language models to meet the specific characteristics of the text sequences.

\subsection{Optimization Loss Analysis (RQ4)}
To evaluate the effectiveness of the contrastive loss, we further conduct comparisons with the BPR loss and report the results in Table~\ref{tab:loss}, from which we can observe better experimental results than BPR loss in our \ac{MNCSR} task.
This result demonstrates the effectiveness of our contrastive loss in helping us model cross-domain information, and thus improve the recommendation performance in all the domains.

\begin{table*}
\centering
\small
\caption{Impact of Different Loss Functions on the \textit{OR-Pantry-Instruments} Dataset.}
\label{tab:loss}
\renewcommand{\arraystretch}{1.3}
\scalebox{1}{
\setlength{\tabcolsep}{1.5mm}{
\begin{tabular}{llcccc}
\toprule

\multicolumn{1}{l}{\textbf{Dataset}} & \multicolumn{1}{l}{\textbf{Loss Type}} & \textbf{Recall@10} & \textbf{NDCG@10} & \textbf{Recall@20} & \textbf{NDCG@20}        \\ \midrule
\multirow{2}{*}{\textit{OR}} & BPR loss & 0.0874  & 0.0440  & 0.1419 & 0.0576    \\
& Contrastive loss & 0.1547 & 0.0734 & 0.2378 & 0.0945 \\
 \cline{2-6}

\multirow{2}{*}{\textit{Pantry}} & BPR loss & 0.0192 & 0.0098  & 0.0277 & 0.0119    \\
& Contrastive loss & 0.0493 & 0.0254 & 0.0709 & 0.0309 \\
 \cline{2-6}

 \multirow{2}{*}{\textit{Instruments}} & BPR loss & 0.0321 & 0.0175 & 0.0473 & 0.0214 \\
& Contrastive loss & 0.0961 & 0.0612 & 0.1239 & 0.0699 \\ \bottomrule

\end{tabular}
}
}
\end{table*}

\subsection{Hyper-parameter Analysis (RQ5)}

\begin{figure*}
    \centering    \includegraphics[width=13cm]{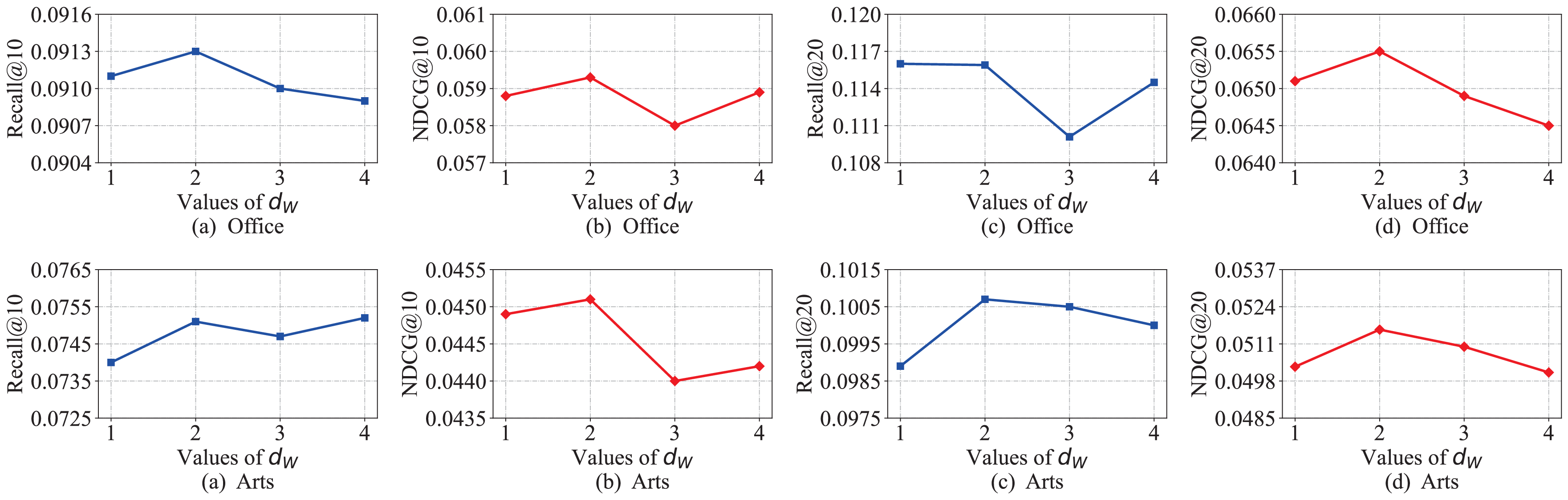}
    \caption{Impact of the hyper-parameter $d_W$ on the \textit{Office-Arts} dataset.}
    \label{fig:hyper_dw}
\end{figure*}

\begin{figure}
    \centering
    \includegraphics[width=8cm]{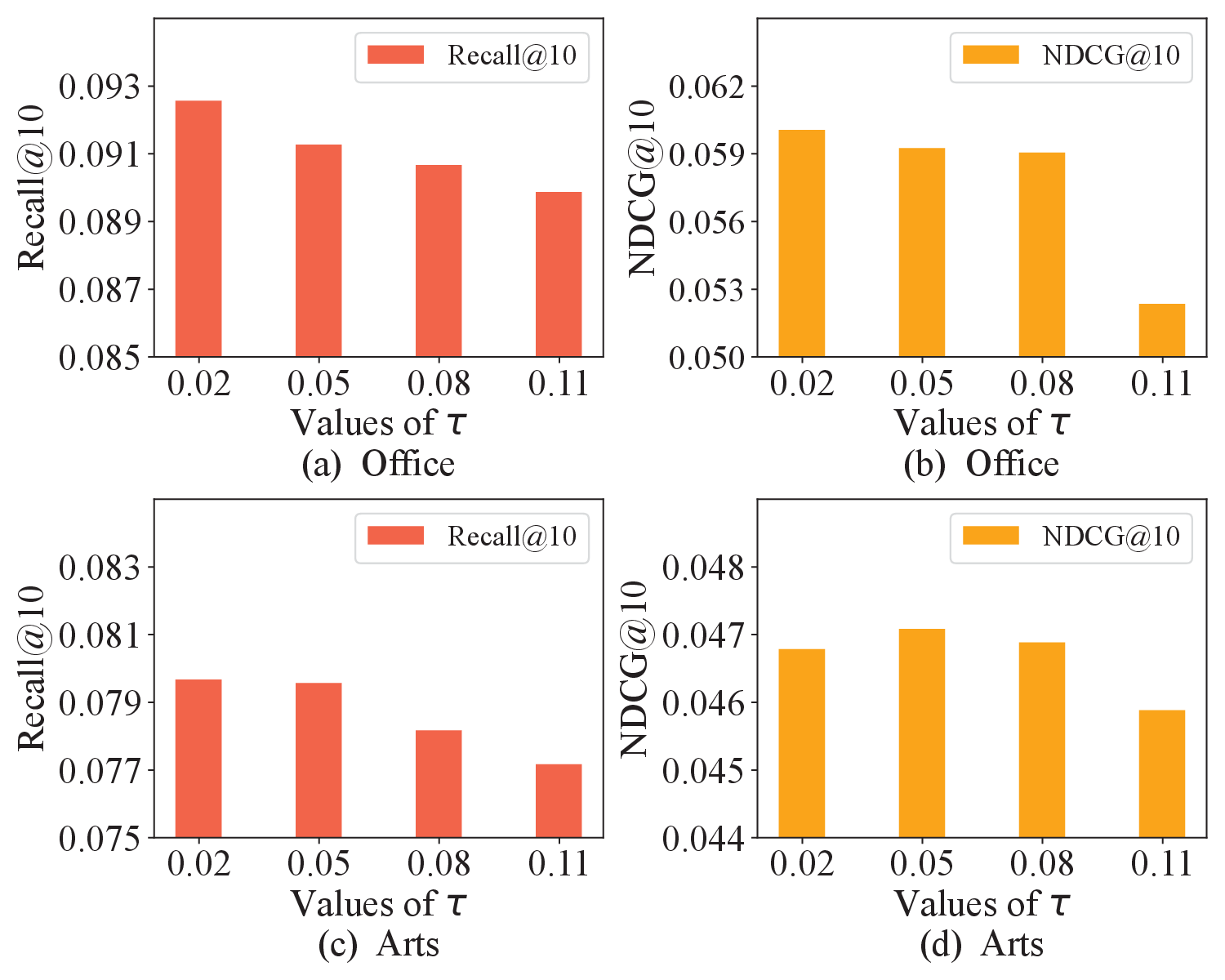}
    \caption{Impact of the hyper-parameter $\tau$ on the \textit{Office-Arts} dataset.}
    \label{fig:hyper_temp}
\end{figure}

We investigate the impact of key hyper-parameters (i.e., the number of prompt contexts $d_W$ and temperature parameter $\tau$) on the performance of \ac{TCPLP} on the \textit{Office-Arts} dataset. 
The hyper-parameter $d_W$ represents the number of context in the prompt, primarily controlling the richness and flexibility of prompts. The hyper-parameter $\tau$ represents the temperature parameter for contrastive learning, serving to adjust the distance between positive and negative samples, enabling the model to better distinguish between them.
The experimental results are respectively shown in Fig.~\ref{fig:hyper_dw} and Fig.~\ref{fig:hyper_temp}.

From Fig.~\ref{fig:hyper_dw}, we can see that a large number of prompt contexts does not necessarily guarantee better performance of \ac{TCPLP}.
Instead, the number of prompt contexts should match the complexity of the knowledge being processed. The best performance of \ac{TCPLP} is observed in the \textit{Office-Arts} dataset when $d_W$ is set to 2.
From Fig.~\ref{fig:hyper_temp}, we can see that the hyper-parameter $\tau$ significantly impacts our method.
When $\tau$ is set to very large values, it leads to the contrastive loss by treating all negative samples equally, making it impossible for our method to distinguish their importance. Conversely, when $\tau$ takes very small values, the \ac{TCPLP} methods mainly focus on those highly challenging negative samples, which may result in difficulties in model convergence and poor generalization ability.
\section{Conclusion}
In this paper, we propose \ac{TCPLP}, a cross-domain sequential recommendation model based on the text-enhanced prompt learning for the \ac{MNCSR} task, where a pre-train \& prompt-tuning paradigm is developed. By leveraging the powerful representational capabilities of language models for sequences and items, we design two types of co-attention-based prompt encoders to capture the prior domain knowledge. To transfer pre-learned knowledge from the source to the target domain, we further exploit the prompt-tuning strategy model adaptation. During the pre-training stage, we devise the domain-shared and domain-specific prompts across multiple domains for prior knowledge learning, while in the prompt-tuning stage, we freeze the domain-shared prompts and only update the domain-specific prompts in the target domain.
We conduct extensive experiments on two subsets of the Amazon dataset (\textit{Office-Arts} and \textit{Cell-Toys-Automotive}) and a cross-platform dataset (\textit{OR-Pantry-Instruments}), and the results demonstrate the effectiveness of our approach compared with the recent SOTA methods. 
Our \ac{TCPLP} method can also be applied to other similar prediction scenarios, such as click-through prediction and rating prediction tasks. The core idea of our method is to establish connections between domains by leveraging the universality of natural language, where a co-attention prompt network is utilized to transfer domain-shared knowledge across multiple domains to the target. This design is independent of the specific task itself and can be applied to other methods using text semantics as the domain connection.


One limitation of this work is that we do not simultaneously consider the atomic Item IDs for representation learning. The atomic IDs of items might also provide additional meaningful information separate from text-based features. Thus, including them could further enrich the item representations and provide more precise recommendations by capturing subtle differences not reflected in text data. We will take solving the above limitation as one of our future works.

\begin{acks}
This work was supported by National Natural Science Foundation of China (Nos. 62372277, 62172263), Natural Science Foundation of Shandong Province (Nos. ZR2022MF257, ZR2020YQ47), Computing Power Internet and Information Security Key Laboratory of Ministry of Education Open Project (2023ZD024),
Youth Innovation Project (No. 2019KJN040), Open Fund CCF-Baidu (No. CCF-BAIDU OF2022008), and Humanities and Social Sciences Fund of the Ministry of Education (No. 21YJC630157).
\end{acks}

\bibliographystyle{ACM-Reference-Format}
\bibliography{Reference}










\end{document}